\documentclass{ws-ijmpa}
\def\GeV{\mbox{GeV}}
\def\TeV{\mbox{TeV}}
\def\Tr{\mbox{Tr}}
\def\Dslash{\rlap{\hspace{0.06cm}/}{D}}

\begin{document}

\markboth{Zhenyu Han}
{Effective Theories and Electroweak Precision Constraints}

%
\catchline{}{}{}{}{}
%

\title{Effective Theories and Electroweak Precision Constraints}

\author{ZHENYU HAN}

\address{Physics Department, University of California, Davis, CA 95616\\
zhenyuhan@physics.ucdavis.edu}

\maketitle

\begin{history}
\received{Day Month Year}
\revised{Day Month Year}
\end{history}

\begin{abstract} 
This is a pedagogical and self-contained review on obtaining electroweak
precision constraints on TeV scale new physics using the effective theory
method. We identify a set of relevant effective operators in the
standard model and calculate from them corrections to all major
electroweak precision observables. The corrections are compared with
data to put constraints on the effective operators. Various approaches
and applications in the literature are reviewed. 

\end{abstract}

\ccode{PACS numbers: 11.60.Cn, 14.80.Bn}

\section{Introduction}
Electroweak precision tests\cite{pdg}\cdash\cite{lectures} (EWPTs)
performed in the last few decades
have achieved great success in establishing the standard model (SM) as
the correct description of electroweak physics. Nevertheless, we are
certain the SM is incomplete and expect new physics to appear at the
TeV scale. The lack of experimental deviations from the SM predictions
does not rule out TeV scale new physics. Rather, it provides us
guidance on the directions we should pursue. The
purpose of this review is to provide readers quick access to
necessary knowledge and
tools for constraining various extensions of the SM. 

In order to be model independent, we adopt the effective field
theory\cite{EFT} (EFT)  
approach. EFT is a powerful tool when there is a distinctive scale 
separating the low energy and the high energy dynamics. The basic
ideal is that the low energy physics can be described by an effective
Lagrangian containing a few degrees of freedom which are not sensitive
to extra degrees of freedom related to the details of the high energy
physics. Fermi's theory of weak interaction is a well known
example. 

For our purpose, given a TeV scale model, we first integrate out all
new physical states and obtain a set of  effective
operators involving the SM fields. These operators are added to
the SM Lagrangian and introduce new interactions among the SM
particles. Consequently they contribute to electroweak observables
(EWPOs) and cause deviations from the SM predictions. Since that the
experiments agree with the SM remarkably well, the deviations must be
small. This is where the constraints come from. The key observation in
favor of the effective theory approach is that the relevant effective
operators are actually much fewer than the vast
possibilities of physics beyond the SM. Therefore, it is natural to
reverse the above procedure, namely, we first calculate
constraints on all possible effective operators (and operator
combinations) without specifying the
model. Once this is done, the remaining work for constraining a given
model is to obtain the operator coefficients in terms of parameters
in that model. In this approach, one needs to make contact with
experimental data only once. In this review, we will show how to calculate 
corrections to EWPOs from the effective operators, 
assuming arbitrary coefficients, while refer readers to
Ref.~\refcite{EFT} on how to obtain  operator coefficients from an underlying
theory.

Schematically, the effective theory can be written as
\begin{equation}
 \label{eq:Lagrangian}
   {\mathcal L}=  {\mathcal L}_{SM} + \sum a_i \, O_i,
 \end{equation}
where ${\mathcal L}_{SM}$ is the standard model Lagrangian and $O_i$'s
are operators containing {\it only} the SM fields. Since we are
interested in extensions of the SM, we assume that
$O_i$'s respect the SM gauge symmetry. Below the new physics scale,
all effects of the new physical states are encoded in $O_i$'s. Of
course one can only work with operators to a given
order and higher order operators have to be truncated. The power
counting depends on whether the electroweak symmetry is linearly or
non-linearly realized. For the former, ${\mathcal L}_{SM}$ is
renormalizable and the electroweak symmetry is broken by the vacuum
expectation value (vev) of a Higgs doublet. The power counting is
determined by the canonical dimensions of the operators $O_i$,
\begin{equation}
{\mathcal L}=  {\mathcal L}_{SM} + \sum a_i \, O_i={\mathcal L}_{SM} +\sum 
\frac{c_i}{\Lambda^{n_i-4}} \, O_i \label{eq:Lagrangian},
\end{equation}
where $n_i$ is the dimension of $O_i$, $c_i$'s are
dimensionless parameters and $\Lambda$ is an energy scale of
$O(\TeV)$. For the non-linear case, the SM spectrum does not contain a
light Higgs boson and ${\mathcal L}_{SM}$ is a non-linear sigma
model. In this review, I will concentrate on the linear case while
also explain how to translate the results to the nonlinear case.

The effective theory breaks down if the new physics
contains physical states much lighter than 1 TeV. For example, the
mass of the lightest neutralino can be as low as $\sim100\GeV$ in the minimal
supersymmetric standard model (MSSM). In this case, there is not an
energy scale that distinctly separates the new physics from the
SM, and an EFT without the new states is not well
defined. Nevertheless, the new states do not appear in initial or
final states in any of the EWPT processes and their dominant
contributions can often be captured in a few effective operators.
 
With the effective Lagrangian (\ref{eq:Lagrangian}) defined, the
theoretical prediction for a given observable $X$ can be written as 
\begin{equation}
X_{th}(a_i)=X_{SM}+\sum a_i X_i,\label{eq:Xth0}
\end{equation} 
where $X_{SM}$ is the SM prediction including loop corrections, which
can be found in the literature\cite{pdg}\cdash\cite{lep2}. $X_i$ is the
correction to $X$
from the operator $O_i$. The current experimental precision allows us
to keep only the corrections linear in the operator coefficients
$a_i$. Most of the review will be devoted to the calculation of $X_i$,
for all well measured EWPOs. 

After obtaining $X_{th}$, we can compare the theoretical predictions with
the measured values of $X$ and get constraints on the coefficients
$a_i$. The constraints are conveniently given in the $\chi^2$
distribution,
\begin{equation}
\label{eq:chi2uncorr} 
  \chi^2(a_i)=\sum_X\frac{\left[X_{th}(a_i)-X_{exp}\right]^2}{\sigma_X^2},
\end{equation} 
where $X_{exp}$ is the experimental value of $X$, and $\sigma_X$
includes both experimental and theoretical errors. In
practice, the above expression is complicated from correlations
between different observables. At this step the advantage of the
effective theory approach is manifest: all relevant experimental
information has been collected in the $\chi^2$
distribution, which
is a simple quadratic function of the operator coefficients $a_i$.
    
There have been many analyses using the effective theory approach
in the literature\cite{Buchmuller+Wyler}\cdash\cite{Han}. The oblique
parameters\cite{oblique}\cdash\cite{WY} are perhaps the best known
example.  The results are often given in
different bases or conventions. In this review, we will adopt the basis in
Ref.~\refcite{Buchmuller+Wyler}, where all independent dimension-6 operators of
the SM are included. They are discussed and trimmed according to our
needs in the next section. In particular, we will identify
the operators relevant to EWPTs.  Section \ref{sec:experiments}
contains a summary of the EWPOs included in this review. In Section
\ref{sec:calculations}, we calculate corrections to EWPOs from the effective
 operators. The constraints are then given in Section
\ref{sec:constraints}. The analysis largely follows 
Ref.~\refcite{Han+Skiba}. Section \ref{sec:discussions} contains a few
discussions, including translations between different bases and conventions.

\section{Effective Operators}\label{sec:operators}
\subsection{The standard model Lagrangian} 
For establishing the notation and later convenience, we give
explicitly the SM Lagrangian,
\begin{eqnarray}
\mathcal{L_{SM}}&=&-\frac14G_{\mu\nu}G^{\mu\nu}-\frac14W_{\mu\nu}W^{\mu\nu}
-\frac14B_{\mu\nu}B^{\mu\nu}\nonumber\\
&&+(D_\mu h)^\dag(D^{\mu}h)+m^2 h^\dag h-\frac12\lambda(h^\dag h)^2\nonumber\\
&&+i\bar{l}{\Dslash}l+i\bar{e}{\Dslash}e+i\bar{q}\Dslash
q+i\bar{u}\Dslash u+i\bar{d}\Dslash d\nonumber\\
&&+(Y_e\bar{l}e h+Y_u\bar{q}u\tilde h+Y_d\bar{q}d h+h.c.),
\end{eqnarray}
where $G_{\mu\nu}$, $W_{\mu\nu}$ and $B_{\mu\nu}$ denote gauge field
strength of the SM gauge groups $SU(3)_c$, $SU(2)_L$ and
$U(1)_Y$, respectively. For simplicity, we have assumed there is only one Higgs
doublet, denoted by $h$. The number of Higgs doublets is irrelevant
for corrections to  EWPOs at tree level,  since all their
contributions are through the electroweak breaking vev,
\begin{equation} 
\langle
h\rangle=\frac{1}{\sqrt{2}}\left(\begin{array}{c}0
\\v\end{array}\right)
\label{eq:hvev}
\end{equation}
with $v=246\GeV$. The left-handed fermion doublets are denoted $q$ and
$l$ and the right-handed singlets $u$, $d$ and $e$. For simplicity, we
have omitted the corresponding subscripts $L$ and $R$ and also the
flavor indices.

Variations of the SM Lagrangian with respect to the fields give us
equations of motion, which we
can use to eliminate redundant operators\cite{Buchmuller+Wyler}. Two
of them will prove useful in our discussion later, which are listed
below. They are obtained by varying $W_\mu^a$ and $B_\mu$ respectively.
\begin{eqnarray}
 (D_\nu
 W^{\nu\mu})^a&=&-\frac12g(ih^\dag\overleftrightarrow D^\mu\sigma^ah+
 \bar l\gamma^\mu \sigma^a l+\bar q \gamma^\mu \sigma^a q),\label{eq:weom}\\
\partial_\nu B^{\nu\mu}&=&-\frac12g'(ih^\dag\overleftrightarrow D^\mu
 h)-g'\sum_fY_f\overline f\gamma^\mu f,\label{eq:beom}
\end{eqnarray}
where $g$ and $g'$ are gauge couplings of $SU(2)_L$ and $U(1)_Y$. $Y_f$
is the hypercharge of fermion $f$.

Substituting the Higgs vev in the SM Lagrangian, we arrive at the mass
eigenstates,
\begin{equation}
Z_\mu=cW_\mu^3-sB_\mu,\quad A_\mu=sW_\mu^3+cB_\mu,\quad W_\mu^\pm=\frac{1}{\sqrt{2}}(W_\mu^1\mp i W_\mu^2),\label{eq:mixing}
\end{equation}
where
\begin{equation}
s=\sin\theta_W=\frac{g'}{\sqrt{g^2+{g'}^2}},\quad
c=\cos\theta_W=\frac{g}{\sqrt{g^2+{g'}^2}}.
\end{equation}
Then the couplings between gauge bosons and fermions can be read from
the covariant derivative,
\begin{eqnarray}
D_\mu&=&\partial_\mu-igW_\mu^aT^a-iYg'B_\mu\nonumber\\
&=&\partial_\mu-i\frac{g}{\sqrt{2}}(W_\mu^+T^{+})-i\frac{e}
{sc}Z_\mu(T^3-s^2Q)-ieA_\mu Q,
\end{eqnarray} 
where $Q$ is the fermion charge and 
\begin{equation}
e=\frac{gg'}{\sqrt{g^2+g'^2}}.
\end{equation}
It is often convenient to write the $Z$-fermion couplings in terms of
$g_V^f$ and $g_A^f$ as defined by
\begin{equation}
\mathcal{L}= \frac{e}{2sc}\overline
f\gamma^\mu(g^f_{V}-g^f_{A}\gamma^5)fZ_\mu.\label{eq:gvga}
\end{equation}
The SM values for $g_V^f$ and $g_A^f$, denoted $g_{V0}^f$ and
$g_{A0}^f$, are given by 
\begin{equation}
\begin{array}{ccc}
  f&g_{V0}^f&g_{A0}^f\\
  \nu_e,\nu_\mu,\nu_\tau&+\frac12&+\frac12\\
  e,\mu, \tau&\  \ -\frac12+2s^2\  \ &\  \ -\frac12\  \ \\
  u,c,t&+\frac12-\frac43s^2&+\frac12\\
  d,s,b&-\frac12+\frac23s^2&-\frac12
\end{array}.\label{eq:gv0ga0}
\end{equation}
\subsection{New operators}
Usually, experiments considered as EWPTs conserve CP, as well as baryon
and lepton numbers. Constraints on processes violating these
symmetries are very stringent, indicating that they are irrelevant to
TeV scale physics. For example, operators contributing to proton decay
have to be suppressed by the grand unification scale or higher. CP
violating operators that contribute to $K-\bar K$ mass
difference or lepton electric dipole moment are typically suppressed by a
scale of $10^3$ to $10^4$ TeV\cite{Buchmuller+Wyler}\,\cite{edm}. 
Therefore, it is convenient to restrict us to
operators conserving CP, and baryon and lepton numbers. 

Similarly, constraints on flavor violating processes are also stringent
(with the exception of processes involving the third
generation). In the lepton sector, the limits on $\mu\rightarrow
e\gamma$ decay implies $10^4$ TeV suppression for contributing
operators\cite{Buchmuller+Wyler}. In the quark sector, limits on new physics
scale obtained from a variety
of $\Delta F=2$ processes are also above $10^3$ to $10^4$ TeV\cite{ut}.
Therefore, we will assume that the operators, and therefore the
observables included in this review all conserve flavors. 

The only dimension-5 operator that respects the SM gauge symmetry is
the neutrino Majorana mass operator, which does not conserve lepton
number and is irrelevant to EWPTs. Not counting flavors, there are 80
independent  dimension-6 operators that conserve
baryon and lepton numbers\cite{Buchmuller+Wyler}, out of which 28
violate CP. We will go through the remaining 52 operators in this section
and identify those contributing to the EWPOs.

For our purpose, it is enough to consider dimension-6 operators
only. Dimension-7 operators do not contribute to EWPTs at tree level, and we do
not need to go to dimension-8 or higher. The reason is the following:
compared with the SM contribution, the corrections
from operator $O_i$ is suppressed by $(v/\Lambda)^{n_i-4}$ or
$(E/\Lambda)^{n_i-4}$ where $E\leq O(200)\GeV$ is the energy scale
involved in the EWPTs. For dimension-6 operators, this suppression is
roughly equal to the experimental precision of the EWPOs when
$\Lambda$ is a few TeV. The suppression factor is much smaller for the
dimension-8 operators which makes their contributions to EWPOs
negligible unless they have extraordinarily large coefficients.

The above reasoning indicates that when calculating the
corrections to the EWPOs from the dimension-6 operators, we only need
to work to the linear order in $a_i=c_i/\Lambda^2$, since the
quadratic corrections are much smaller and therefore can be
neglected. This means we only need to work at tree level for those
operators\footnote{The exception is muon $(g-2)$ discussed later. Due
  to the high precision of this measurement, loop level diagrams are
  also important.}. Moreover, when calculating from the matrix
element to the cross-section, only the
{\it interferences} between the SM Feynman diagrams and diagrams induced
by new operators are important. This important observation will be used
to reduce the number of operators.

The linear-order assumption also helps us understand why we only need
the vev of the Higgs doublet: the physical Higgs boson contributes to
EWPOs only at loop levels, any corrections to the Higgs mass or
coupling from  dimension-6 operators will be suppressed by an extra
loop factor when calculating the corrections to EWPOs and therefore negligible.

The 52 dimension-6 operators that conserve CP are enumerated
below. The naming scheme follows Ref.~\refcite{Buchmuller+Wyler}, with
small changes in the notation.  We will
identify those that are relevant to EWPTs by general arguments
in this section. More detailed calculations are presented in 
Sec.~\ref{sec:calculations}. If the analysis here is too
sketchy for the reader, it should become clear later on. 

\begin{enumerate}
\item Vectors only
\begin{eqnarray}
O_G&=&\epsilon^{abc}  G^{a \nu}_{\mu} G^{b\lambda}_{\nu} G^{c
  \mu}_{\lambda},\nonumber\\
O_W &=& \epsilon^{abc} W^{a \nu}_{\mu} W^{b\lambda}_{\nu} W^{c \mu}_{\lambda}.
\end{eqnarray}
The two operators affect triple gauge couplings. At tree level, the
former only affects pure hadronic processes, which are not measured as
well as EWPOs.
Therefore, only the operator $O_W$ is interesting to us.
\item Fermions only
\begin{enumerate}
\item $\overline L L\overline L L$ operators
\begin{eqnarray} &&
  O_{ll}^{(1)}=\frac{1}{2} (\overline{l} \gamma^\mu l) (\overline{l}
  \gamma_\mu l), 
\ \ \
  O_{ll}^{(3)}=\frac{1}{2} (\overline{l} \gamma^\mu \sigma^a l) 
(\overline{l} \gamma_\mu \sigma^a l),
      \label{eq:oll} \nonumber\\ &&
  O_{qq}^{(1,1)}=\frac{1}{2} (\overline{q} \gamma^\mu q) (\overline{q}
  \gamma_\mu q), 
\ \ \
  O_{qq}^{(8,1)}=\frac{1}{2} (\overline{q} \gamma^\mu\lambda^A  q) 
(\overline{q} \gamma_\mu\lambda^A q),
       \nonumber\\ &&
  O_{qq}^{(1,3)}=\frac{1}{2} (\overline{q} \gamma^\mu \sigma^a q) (\overline{q}
  \gamma_\mu\sigma^a q), 
\ \ \
  O_{qq}^{(8,3)}=\frac{1}{2} (\overline{q} \gamma^\mu\lambda^A \sigma^a q) 
(\overline{q} \gamma_\mu\lambda^A \sigma^a q),
      \nonumber \\ &&
 O_{lq}^{(1)}=\frac{1}{2} (\overline{l} \gamma^\mu l) (\overline{q}
  \gamma_\mu q), 
\ \ \
  O_{lq}^{(3)}=\frac{1}{2} (\overline{l} \gamma^\mu \sigma^a l) 
(\overline{q} \gamma_\mu \sigma^a q). 
\end{eqnarray}
\item $\overline R R\overline R R$ operators
 \begin{eqnarray}
 &&  O_{ee}=\frac{1}{2} (\overline{e} \gamma^\mu e) (\overline{e}
  \gamma_\mu e), \nonumber\\
&&O_{uu}^{(1)}=\frac{1}{2} (\overline{u} \gamma^\mu u) (\overline{u}
  \gamma_\mu u), \ \ \ O_{uu}^{(8)}=\frac{1}{2} (\overline{u} \lambda^A\gamma^\mu u) (\overline{u}\lambda^A
  \gamma_\mu u),\nonumber\\
&&O_{dd}^{(1)}=\frac{1}{2} (\overline{d} \gamma^\mu d) (\overline{d}
  \gamma_\mu d), \ \ \ O_{dd}^{(8)}=\frac{1}{2} (\overline{d} \lambda^A\gamma^\mu d) (\overline{d}\lambda^A
  \gamma_\mu d),\nonumber\\
  && O_{eu}=(\overline{e} \gamma^\mu e) (\overline{u} \gamma_\mu u), \nonumber \\
  && O_{ed}=(\overline{e} \gamma^\mu e) (\overline{d} \gamma_\mu d),\nonumber \\
 && O_{ud}^{(1)}=\frac{1}{2} (\overline{u} \gamma^\mu u) (\overline{d}
  \gamma_\mu d), \ \ \ O_{ud}^{(8)}=\frac{1}{2} (\overline{u} \lambda^A\gamma^\mu u) (\overline{d}\lambda^A
  \gamma_\mu d).
\end{eqnarray}
\item $\overline L L\overline R R$ operators\\
For convenience, we have changed the form from $\overline L R\overline R L$
in Ref.~\refcite{Buchmuller+Wyler} to $\overline L L\overline R R$ by
using the Fierz transformation
\begin{equation*}
(\overline{\psi}_{1L}\psi_{2R})(\overline{\psi}_{3R}\psi_{4L})=\frac12(\overline{\psi}_{1L}\gamma^
\mu\psi_{4L})(\overline{\psi}_{3R}\gamma_\mu\psi_{2R}).
\end{equation*}
\begin{eqnarray}
 && O_{le}= (\overline{l} \gamma^\mu l) (\overline{e} \gamma_\mu e),\nonumber \\
 && O_{lu}= (\overline{l} \gamma^\mu l) (\overline{u} \gamma_\mu u),\nonumber\\
 && O_{ld}= (\overline{l} \gamma^\mu l) (\overline{d} \gamma_\mu d),\nonumber\\
 && O_{qe}=(\overline{q} \gamma^\mu q) (\overline{e} \gamma_\mu e),\nonumber\\
 && O_{qu}^{(1)}=(\overline{q} \gamma^\mu q) (\overline{u} \gamma_\mu u),
   \ \ \ O_{qu}^{(8)}=(\overline{q} \gamma^\mu\lambda^A q)
 (\overline{u} \gamma_\mu\lambda^A u),\nonumber\\
 && O_{qd}^{(1)}=(\overline{q} \gamma^\mu q) (\overline{d} \gamma_\mu d),
   \ \ \ O_{qd}^{(8)}=(\overline{q} \gamma^\mu\lambda^A q)
 (\overline{d} \gamma_\mu\lambda^A d),\nonumber\\
&&O_{qde}=(\overline{l}\gamma^\mu q)(\overline{d} \gamma_{\mu} e).
\end{eqnarray}
\item $\overline L R\overline L R$ operators
\begin{eqnarray}
  &&O_{qq}^{(1)}=\epsilon^{ab}(\overline{q}^a u) (\overline{q}^b d),\ \ \
  O_{qq}^{(8)}=\epsilon^{ab}(\overline{q}^a\lambda^A u) (\overline{q}^a\lambda^A d),\nonumber\\
&&O_{lq}=\epsilon^{ab}(\overline{l}^a e) (\overline{q}^b u).
\end{eqnarray}
\end{enumerate}
Again we can safely ignore those operators containing four
quarks. For the operators $O_{qde}$ and $O_{lq}$, they contain
pieces $\overline e_L e_R\overline d_R d_L$ and $\overline e_L
e_R\overline u_L d_R$ that contribute to the process
$e^+e^-\rightarrow \mbox{hadrons} $. However, the corresponding SM
contribution are highly suppressed by electron and quark Yukawas,
which are negligible. Since we are only interested in interferences
between the SM and the new physics, the contributions from $O_{qde}$ and
$O_{lq}$ are negligible too. Thus the interesting operators are:
$O_{ll}^{(1)}$, $O_{ll}^{(3)}$, $O_{lq}^{(1)}$, $O_{lq}^{(3)}$,
$O_{ee}$, $O_{eu}$, $O_{ed}$, $O_{le}$, $O_{lu}$, $O_{ld}$, $O_{qe}$.
\item Scalars only
\begin{eqnarray}
&&O_h=\frac13(h^\dag h)^3,\nonumber\\
&&O_{\partial h}=\frac12\partial_\mu(h^\dag h)\partial^\mu(h^\dag h).
\end{eqnarray}
The two operators affect the Higgs mass and self-coupling, which are
negligible as explained earlier. 
\item Fermions and vectors
\begin{eqnarray}
&&O_{lW}=i\overline l\sigma^a\gamma^\mu D^{\nu}lW^{a}_{\mu\nu},\ \ \ O_{lB}=i\overline l\gamma^\mu D^\nu lB_{\mu\nu},\nonumber\\
&&O_{eB'}=i\overline e \gamma^\mu D^\nu eB_{\mu\nu},\nonumber\\
&&O_{qG}=i\overline q \lambda^A\gamma^\mu D^\nu qG_{A\mu\nu},\nonumber\\
&&O_{qW}=i\overline q \sigma^a\gamma^\mu D^\nu qW^a_{\mu\nu},\ \ \ O_{qB}=i\overline q\gamma^\mu D^\nu qB_{\mu\nu},\nonumber\\
&&O_{uG}=i\overline u\lambda^A\gamma^\mu D^\nu uG^A_{\mu\nu},\nonumber\\
&&O_{uB}=i\overline u \gamma^\mu D^\nu uB_{\mu\nu},\nonumber\\
&&O_{dG}=i\overline d\lambda^A\gamma^\mu D^\nu dG^A_{\mu\nu},\nonumber\\
&&O_{dB}=i\overline d \gamma^\mu D^\nu dB_{\mu\nu}.
\end{eqnarray}
At first sight, these operators alter couplings between the gauge
bosons and the fermions. However, the couplings are imaginary compared
with the SM couplings. Therefore the SM diagrams do not interfere with
the new diagrams.\footnote{An exception is, at the $Z$-pole, the
$Z$-exchange diagrams with the new couplings do interfere with the SM
$\gamma$-exchange diagrams and vice versa. However, these interferences
are  suppressed further by $\Gamma_Z/M_Z$ besides the factor
$v^2/\Lambda^2$. Therefore, they are negligible too.} In this case,
corrections from these operators can be ignored.
\item Scalars and vectors
\begin{eqnarray}
&&O_{hG}=\frac12(h^\dag h)G_{\mu\nu}^AG^{A\mu\nu},\nonumber\\
&&O_{hW}=\frac12(h^\dag h)W_{\mu\nu}^aW^{a\mu\nu},\nonumber\\
&&O_{hB}=\frac12(h^\dag h)B_{\mu\nu}B^{\mu\nu},\nonumber\\
&&O_{WB}=(h^\dag\sigma^a h)W^a_{\mu\nu}B^{\mu\nu},\nonumber\\
&&O_h^{(1)}=(h^\dag h)(D_\mu h^\dag D^\mu h),\ \ \ O_h^{(3)}=(O_h)=(h^\dag D^\mu h)(D_\mu h^\dag h).
\end{eqnarray}
When $h$ acquires a vev, the operators
$O_{hG}$, $O_{hW}$, $O_{hB}$, $O_h^{(1)}$ yield corrections to the
kinetic terms for $G_{\mu\nu}$, $W_{\mu\nu}$, $B_{\mu\nu}$ and $h$,
which can be absorbed by field redefinitions. Therefore, they do not
contribute to EWPOs at the order $1/\Lambda^2$. The other two
operators $O_{WB}$ and $O_h^{(3)}$ (simplified as $O_h$ in the rest of
the review) modify the gauge boson
propagators. Indeed, they correspond
to the well-known oblique $S$ and $T$ parameters. One also notices
that $O_{WB}$ contains corrections to triple gauge couplings. 
\item Fermions and scalars
\begin{eqnarray}
&&O_{eh}=(h^\dag h) (\overline l e h),\nonumber\\
&&O_{uh}=(h^\dag h) (\overline q u \tilde h),\nonumber\\
&&O_{eh}=(h^\dag h) (\overline q d h).
\end{eqnarray}
The effects of these operators can be absorbed in the SM Yukawa
couplings and therefore do not make any contribution.
\item Fermions, gauge bosons and scalar
\begin{enumerate}
\item One derivative
\begin{eqnarray}
&&O_{hl}^{(1)}=i(h^\dag D_\mu h)(\overline l\gamma^\mu l)+h.c.,\nonumber\\
&&O_{hl}^{(3)}=i(h^\dag D_\mu\sigma^a h)(\overline l\gamma^\mu\sigma^a
  l)+h.c.,\nonumber\\
&&O_{he}=i(h^\dag D_\mu h)(\overline e\gamma^\mu e)+h.c.,\nonumber\\
&&O_{hq}^{(1)}=i(h^\dag D_\mu h)(\overline q\gamma^\mu qq)+h.c.,\nonumber\\
&&O_{hq}^{(3)}=i(h^\dag D_\mu\sigma^a h)(\overline q\gamma^\mu\sigma^a
  q)+h.c.,\nonumber\\
&&O_{hu}=i(h^\dag D_\mu h)(\overline u\gamma^\mu u)+h.c.,\nonumber\\
&&O_{hd}=i(h^\dag D_\mu h)(\overline d\gamma^\mu d)+h.c.,\nonumber\\
&&O_{hh}=i( h^T \epsilon D_\mu h)(\overline u \gamma^\mu d)+h.c..\label{eq:hff}
\end{eqnarray}
When $h$ gets a vev, these operators modify the gauge-fermion
couplings and therefore are important for our calculation. It is worth
noting that the operator $O_{hh}$ induces a coupling in the form
$W^+_\mu\overline u_R\gamma^\mu d_R+h.c.$ which is absent in the SM. In other words,
there is no interference for diagram containing this coupling and the
SM diagram. Therefore, we expect loose bound on the operator
$O_{hh}$.
\item Two derivatives
\begin{eqnarray}
&&O_{De}=(\overline l D_\mu e)D^\mu h,\ \ \
 O_{\overline De}=(D_\mu\overline l e)D^\mu h,\nonumber\\
&&O_{Du}=(\overline q D_\mu u)D^\mu \tilde h,
\ \ \ O_{\overline Du}=(D_\mu\overline q e)D^\mu \tilde h,\nonumber\\
&&O_{Dd}=(\overline q D_\mu d)D^\mu h,
\ \ \ O_{\overline Dd}=(D_\mu\overline q d)D^\mu \tilde h,\nonumber\\
&&O_{eW}=(\overline l\sigma^{\mu\nu}\sigma^a e)hW^a_{\mu\nu},\ \ \
  O_{eB}=(\overline l\sigma^{\mu\nu} e)h B_{\mu\nu},\nonumber\\
&&O_{uG}=(\bar\sigma^{\mu\nu}\lambda^A u)\tilde h G_{\mu\nu}^A,\nonumber\\
&&O_{uW}=(\overline q\bar\sigma^{\mu\nu}\sigma^a u)\tilde h W_{\mu\nu}^a,\ \ \ 
O_{uB}=(\overline q\bar\sigma^{\mu\nu} u)\tilde h B_{\mu\nu},\nonumber\\
&&O_{dG}=(\overline q\bar\sigma^{\mu\nu}\lambda^A u)h G_{\mu\nu}^A,\nonumber\\
&&O_{uW}=(\overline q\bar\sigma^{\mu\nu}\sigma^a u) h W_{\mu\nu}^a,\ \ \ 
O_{uB}=(\overline q\bar\sigma^{\mu\nu} u) h B_{\mu\nu}.
\end{eqnarray}
Like operators in Eqs.~(\ref{eq:hff}), Hermitian conjugate should be
added to each operator above. Most of these operators do not have a
tight bound from EWPTs. They
all produce couplings absent in the SM in the tree level and therefore
do not interfere with the SM diagrams. The exception is $O_{eW}$ and
$O_{eB}$, which contribute at tree level to the lepton
$(g-2)$. Indeed, the precision for electron and muon $(g-2)$
measurements are so good that if we naively take the operator
coefficients to be $O(1/\Lambda^2)$, the bound on $\Lambda$ will be
larger than $100$ TeV. Usually, the contribution from TeV scale
physics is suppressed by both the loop factor and the lepton mass, which
makes it too small to compare with the experiments for the electron
$g-2$\footnote{Electron $g-2$ is one of the measurements from which the
  input parameter $\alpha$ is determined.} and the bounds from muon
$g-2$ comparable to the other
EWPOs. Other dimension-6 operators can also make sizable contributions
to muon $(g-2)$ at one loop level. We will come back to this observable later.   
\end{enumerate}
\end{enumerate}
The relevant operators are summarized below. Following
Ref.~\refcite{Han+Skiba}, we have simplified the notation by replacing
superscripts $(1)$ and $(3)$ with $s$
and $t$, denoting singlets and triplets respectively.
\begin{enumerate}
\item Operators that modify gauge boson propagators
\begin{equation}
O_{WB},O_h.\label{eq:owb}
\end{equation}
\item Operators that affect tree level SM gauge-fermion couplings
\begin{eqnarray}
&O_{hl}^s,O_{hl}^t,O_{he},O_{hq}^s,O_{hq}^t,O_{hu},O_{hd}.\label{eq:ohf}
\end{eqnarray}
\item Four-fermion opertors
\begin{equation}
O_{ll}^s,O_{ll}^t,O_{lq}^s,O_{lq}^t,O_{le},O_{qe},O_{lu},O_{ld},O_{ee},O_{eu},
O_{ed}.\label{eq:4fermi} 
\end{equation}
\item Operators that modify triple gauge boson couplings
\begin{equation}
O_W, (O_{WB}).\label{eq:ow}
\end{equation}
\item Operators that modify lepton $g-2$
\begin{equation}
O_{eW},O_{eB}.\label{eq:oew}
\end{equation}
\end{enumerate}

We have not specified the flavor structure for the above operators
except the assumption that they conserve flavors. Flavors can be
conserved separately for each generation. In this case, each of the
$O_i$'s actually corresponds to several operators with different
flavors and independent coefficients. This is the most general case.
However, it requires the quarks in the operators to be mass
eigenstates. Therefore the new physics must be aligned with the Yukawa
matrices, which seems unnatural and is difficult to realize. A simpler
option is imposing flavor universality for the effective
operators. This is natural for flavor-blind new physics, for example,
a $Z'$ gauge boson that couples to all generations universally. In
this case, each $O_i$ corresponds to several operators with different
flavors but the same coefficient $a_i$. The operators should be
understood to be contracted over flavor indices (in the 4-fermion
operators, flavor indices are contracted over fermions (12) and (34)
respectively. In the following analysis, unless specified, we will assume
flavor universality. Nevertheless, calculations for flavor universal and
non-universal cases are very similar and it is straightforward
to translate from one case to the other.

Flavor universality is assumed in the analysis of
Ref.~\refcite{Han+Skiba}. It is realized through a global $U(3)^5$
symmetry, with each $U(3)$ corresponds respectively to each of $q$,
$l$, $u$, $d$ and $e$. Under the $U(3)$, fermions and antifermions
transform as fundamental and anti-fundamental representations. This is
the largest symmetry of the fermion kinetic
terms in the SM, i.e., the symmetry when the Yukawa couplings are
turned off. This
symmetry not only guarantees flavor universality, but also forbids some
of the operators discussed above, for example,
$O_{qde}=(\overline{l}\gamma^\mu q)(\overline{d} \gamma_{\mu}
e)$, because left-handed fields and right-handed fields transform
under different $U(3)$'s. Most of those operators do not
interfere with the SM, and therefore have already been excluded from
the list Eq.~(\ref{eq:owb})-(\ref{eq:oew}). The only exception is the
operators $O_{eB}$ and $O_{eW}$, which contribute to the muon
$g-2$, but is excluded in Ref.~\refcite{Han+Skiba}.  In this review, we
will quote results from
Ref.~\refcite{Han+Skiba} whenever is appropriate, and also include
discussions on the muon $g-2$. An alternative flavor symmetry is
discussed in Section \ref{sec:discussions}.

Following a summary of the electroweak observables in the next
section, we will
consider one by one the operators in Eqs.~(\ref{eq:owb})-(\ref{eq:oew})
and calculate their corrections to the EWPOs in Section \ref{sec:calculations}. 

\section{Experiments}
\label{sec:experiments}
As mentioned before, measurements included in this review do not
contain those violating CP or flavor symmetries. The observables,
together with their experimental values and SM predictions are
summarized in Table \ref{table:experiments}. The experimental values
and the SM predictions are excerpted from Ref.~\refcite{pdg} except
LEP 2 measurements.  Multiple
measurements for the same observable have been combined.
We will give more information for each observable when calculating 
corrections from the effective operators. The information included in this
review is minimal for understanding the calculations. It is
strongly recommended that readers 
consult more comprehensive reviews for further
details\cite{pdg}\cdash\cite{lep2}. 
\begin{table}[h]
\tbl{ Relevant measurements}
{\begin{tabular}{cccc}
\hline
 & Notation & Value & SM prediction \\
 \hline
 Atomic parity &$Q_W(Cs)$&$-72.62\pm0.46$&$-73.17\pm0.03$\\
 violation& $Q_W(Tl)$& $-116.6\pm3.7$&$-116.78\pm0.05$\\
 \hline
 Muon $g-2$&$\frac12(g_\mu-2-\frac{\alpha}{\pi})[10^{-9}]$&$4511.07\pm0.82$&$4509.82\pm0.10$\\
\hline
   $\nu$-nucleon scattering&$g_L^2$&$0.30005\pm0.00137$&$0.30378\pm0.0002$1\\
       &$g_R^2$&$0.03076\pm0.00110$&$0.03006\pm0.00003$\\
 \hline   
   $\nu$-$e$ scattering &$g_V^{\nu e}$&$-0.040\pm0.015$&$-0.0396\pm0.0003$\\
    &$g_A^{\nu e}$&$-0.507\pm0.014$&$-0.5064\pm0.0001$\\
 \hline
   $e^+e^-\rightarrow f\bar f$&$\Gamma_Z[\GeV]$&$2.4952\pm0.0023$&$2.4968\pm0.0011$\\
    at $Z$-pole &$\sigma_h^0[\mbox{nb}]$ &$41.541\pm0.037$&$41.467\pm0.009$\\
         &$R_e^0$&$20.804\pm0.050$ &$20.756\pm0.011$\\
         &$R_\mu^0$&$20.785\pm0.033$&$20.756\pm0.011$\\
         &$R_\tau^0$&$20.764\pm0.045$&$20.801\pm0.011$\\
         &$R_b$&$0.21629\pm0.00066$&$0.21578\pm0.00010$\\
         &$R_c$&$0.1721\pm0.0030$&$0.17230\pm0.00004$\\
         &$A_{fb}^{0,e}$&$0.0145\pm0.0025$&$0.01622\pm0.00025$\\
         &$A_{fb}^{0,\mu}$&$0.0169\pm0.0025$&$0.01622\pm0.00025$\\
         &$A_{fb}^{0,\tau}$&$0.0188\pm0.0017$&$0.01622\pm0.00025$\\
         &$A_{fb}^{0,b}$&$0.0992\pm0.0016$&$0.1031\pm0.0008$\\
         &$A_{fb}^{0,c}$&$0.0707\pm0.0035$&$0.0737\pm0.0006$\\
      &$\sin^2\theta_{eff}^{lept}(Q_{fb})$&$0.2319\pm0.0012$&$0.23152\pm0.00014$\\
         &$A_e$&$0.1514\pm0.0019$&$0.1471\pm0.0011$\\
         &$A_\mu$&$0.142\pm0.015$&$0.1471\pm0.0011$\\
          &$A_\tau$&$0.1433\pm0.0041$&$0.1471\pm0.0011$\\
 \hline
  Fermion pair
  &$\sigma_f(f=q,\mu,\tau)$&Ref.~\refcite{lep2}&Ref.~\refcite{lep2}\\ production
  at
  &$A_{fb}^f(f=\mu,\tau)$&Ref.~\refcite{lep2}&Ref.~\refcite{lep2}\\ LEP
  2            &$d\sigma_e/d\cos\theta$&Ref.~\refcite{OPALfpair}&Ref.~\refcite{BHWIDE}\\ \hline
 $W$ pair&$d\sigma_W/d\cos\theta$&Ref.~\refcite{L3Wpair} &Ref.~\refcite{L3Wpair}\\
 \hline
 $W$ mass&$M_W[\GeV]$ &$80.410\pm0.032$  &$80.376\pm0.017$\\
 \hline
\end{tabular}\label{table:experiments}}
\end{table}

Not shown in Table \ref{table:experiments} are three most precisely
measured observables: the fine structure constant $\alpha$, Fermi
constant $G_F$ determined from muon lifetime and the $Z$ boson mass
measured at LEP 1. The experimental values are\cite{pdg}
\begin{eqnarray}
&\alpha=1/137.03599911(46),\ \ G_F=1.16637(1)\times
10^{-5}\GeV^{-2},\nonumber\\& M_{Z}=\,91.1876\pm0.0021\GeV.
\end{eqnarray}  
 Not counting Higgs and fermion masses and mixings, the SM Lagrangian has
three parameters $g$, $g'$ and $v$.  However,
they are not directly measurable. Therefore we take the above three
observables as the input parameters, from which the SM predictions are
calculated. The SM values also depend on the top quark mass $m_t$,
the Higgs mass $m_h$, and the strong coupling constant
$\alpha_s$, for which the best fit values are\cite{pdg}
\begin{equation}
m_t=172.7\pm 2.8\GeV,\quad m_H=89^{+38}_{-28}\GeV, \quad
\alpha_s(M_Z)=0.1216\pm 0.0017.
\end{equation}
It is interesting that the data favors a light Higgs, with an upper bound
of $\sim200\GeV$ for the mass at 95\%
confidence level (CL). 

The SM predictions with the above input parameters agree with the
experimental values very well. The $\chi^2$ per degree of freedom is
1.1 (excluding LEP 2 measurements), and only two
observables have deviations exceeding 2 $\sigma$, -2.4 for
$A_{fb}^{0,b}$ and -2.7 for $g_L^2$. 

Note that the observables listed in Table \ref{table:experiments} are
often ``pseudo-observables'', in the sense that they are not the directly
measured quantities in the experiments. For example, the LEP 1 experiments
measured the cross sections for $e^+e^-\rightarrow f\overline f$ at a
few different center mass energy around the $Z$-pole. They have all
been combined and translated to a few quantities at $Z$-pole. For LEP
2 measurements at various center of mass energies, there is no such
simplification and therefore the numerical values of SM predictions
are not listed in Table \ref{table:experiments}. Instead, we have
given the corresponding references. 

The $Z$-pole observables and several low-energy observables have the
best precision. They dominate the constraints whenever the
considered operators contribute to them. Therefore, in the literature,
some of the observables in Table \ref{table:experiments} are often
omitted, which does not significantly alter the bounds on the models in
consideration. Nevertheless, there are also
operators that cannot be constrained by the $Z$-pole and low energy
measurements. As we will
see, some of the 4-fermion operators are only constrained by the LEP 2
measurements. The LEP 2 measurement for $e^+e^-\rightarrow W^+W^-$
cross sections also provides us unique constraints on triple gauge
couplings.    

\section{Calculations}
\label{sec:calculations}
\subsection{General consideration}
We now start to calculate the corrections from operators (\ref{eq:owb})
through (\ref{eq:oew}) to the
EWPOs listed in Table \ref{table:experiments}. In other words, we want to
calculate the values of $X_i$ in Eq.~(\ref{eq:Xth0}), which is the
leading order contribution to observable $X$ from the operator
$O_i$. Before obtaining $X_i$ for each observable $X$, let us consider
in more detail how those operators contribute:
\begin{enumerate}
\item Operators that modify gauge boson propagators
\begin{enumerate}
\item $O_{WB}$\\
Substituting in the vev of $h$, we have
\begin{eqnarray}
    &&a_{WB}h^\dag W^{a\mu\nu}\sigma^ah B_{\mu\nu}\nonumber\\
    &=&-a_{WB}\frac{v^2}2[sc(A^{\mu\nu}A_{\mu\nu}-Z^{\mu\nu}Z_{\mu\nu})+(c^2-s^2)A^{\mu\nu}Z_{\mu\nu}]\nonumber\\
&&-a_{WB}v^2gf^{abc}W^{b\mu}W^{c\nu}\partial_\mu B_\nu,\label{eq:owb_exp}   
\end{eqnarray}
where we have used Eq.~(\ref{eq:mixing}).
From Eq.~(\ref{eq:owb_exp}), we can identify the corrections to the gauge
boson propagators
\begin{equation}
 \Pi'_{Z\gamma}=-a_{WB}v^2(c^2-s^2),\quad
\Pi'_{\gamma\gamma}=-2a_{WB}v^2sc,\quad\Pi'_{ZZ}=2a_{WB}v^2sc.\label{eq:piprime}
\end{equation}
After field redefinitions 
\begin{equation}
Z_\mu\rightarrow Z_\mu(1-\Pi'_{ZZ})^{\frac12},\ \ 
A_\mu\rightarrow A_\mu(1-\Pi'_{\gamma\gamma})^{\frac12},\label{eq:zaredef}
\end{equation}
the kinetic terms become canonical and the physical $Z$ mass is
shifted to 
\begin{equation}
M_Z^2= M_{Z0}^2 (1+\Pi'_{ZZ})= M_{Z0}^2(1+2a_{WB}v^2sc),
\end{equation}
where $M_{Z0}=\sqrt{g^2+g'^2}v/2$.
Another effect of the field redefinition is that the couplings between
the physical gauge bosons, $\gamma$ and $Z$, and the fermions are
modified by a factor
$(1+\Pi'_{\gamma\gamma})^{1/2}$ and $(1+\Pi'_{ZZ})^{1/2}$ respectively. In
particular, the fine structure constant becomes
\begin{equation}
\alpha=\frac{e^2}{4\pi}(1+\Pi'_{\gamma\gamma})=\frac{e^2}{4\pi}(1-2a_{WB}v^2sc).
\end{equation}  
The mixing term $\Pi'_{\gamma Z}$ makes another correction to the
$Z$-fermion couplings in the form 
\begin{equation}
\Delta g_V^f=2sc\Pi'_{Z\gamma}Q_f,
\end{equation}
where $Q_f$ is the electric charge of the fermion $f$.

\item $O_h$\\
\begin{equation}
a_h|h^\dag D_\mu h|^2=a_h\frac{v^4}{16}\frac{g^2}{c^2}Z_\mu Z^\mu.\label{eq:oh_exp}
\end{equation}
This is a direct contribution to the $Z$ mass:
\begin{equation}
\Delta M_Z^2=\Pi_{ZZ}(0)=a_h\frac{v^2}{8}\frac{g^2}{c^2}.\label{eq:pizz0}
\end{equation}
\end{enumerate}
\item Operators that modify gauge-fermion couplings\\
These operators all contain 2 Higgs doublets and 2 fermions and
a covariant derivative which affect the $Z$-fermion
couplings $g_V^f$ and $g_A^f$ defined in Eq.~(\ref{eq:gvga}).
For example, from 
\begin{eqnarray}
a_{hl}^si(h^\dag D_\mu h)(\overline l\gamma^\mu l)+h.c. 
   &=& \frac{gv^2}{2c}Z_\mu
(\overline\nu\gamma^\mu\nu+
\overline{e}_{L}\gamma^\mu e_{L}),
\end{eqnarray}
we read
\begin{equation}
\Delta g_V^{e}=\Delta
g_A^{e}=\Delta g_V^{\nu}=\Delta
g_A^{\nu}=-a_{hl}^s\frac{v^2}{2}.
\end{equation}
The corrections from the other operators are similar. Combining all
contributions, we obtain
\begin{eqnarray}
 \Delta g_V^e&=&\Pi_{Z\gamma}'Q_e2sc-\frac{v^2}{2}a^s_{hl}-\frac{v^2}{2}a_{hl}^t
 -\frac{v^2}2a_{he},\nonumber\\
 \Delta g_A^e&=&-\frac{v^2}{2}a_{hl}^s-\frac{v^2}{2}a_{hl}^t+\frac{v^2}2a_{he},\nonumber\\
 \Delta g_V^\nu&=&-\frac{v^2}{2}a_{hl}^s+\frac{v^2}{2}a_{hl}^t,\nonumber\\
 \Delta g_A^\nu&=&-\frac{v^2}{2}a_{hl}^s+\frac{v^2}{2}a_{hl}^t,\nonumber\\
 \Delta g_V^u&=&\Pi_{Z\gamma}'Q_u
 2sc-\frac{v^2}{2}a_{hq}^s+\frac{v^2}{2}a_{hq}^t -\frac{v^2}2a_{hu},\nonumber\\
 \Delta g_A^u&=&-\frac{v^2}{2}a_{hq}^s+\frac{v^2}{2}a_{hq}^t+\frac{v^2}2a_{hu},\nonumber\\
 \Delta g_V^d&=&\Pi_{Z\gamma}'Q_d 2sc-\frac{v^2}{2}a_{hq}^s-\frac{v^2}{2}a_{hq}^t-\frac{v^2}2a_{hd},\nonumber\\
 \Delta g_A^u&=&-\frac{v^2}{2}a_{hq}^s-\frac{v^2}{2}a_{hq}^t+\frac{v^2}2a_{hu},\label{eq:dgvdga}
\end{eqnarray}
where we have added the contributions from $\Pi'_{Z\gamma}$ given in
(\ref{eq:piprime}). As mentioned, there is also a contribution from
the field redefinition of $Z$ for each coupling,
\begin{equation}
\Delta g_{V,A}^f=\frac12g_{V0,A0}^f\Pi'_{ZZ}=g_{V0,A0}^fa_{WB}scv^2.
\end{equation}

The operators $a_{hl}^t$ and
$a_{hq}^t$ also modify the  $W$-fermion couplings:
\begin{eqnarray}
g&\rightarrow& g\left(1+v^2a_{hl}^{t}\right), \quad(W-\mbox{lepton
  couplings}),\nonumber \\
g&\rightarrow& g\left(1+v^2a_{hq}^{t}\right),\quad (W-\mbox{quark
  couplings}). \label{eq:wff} 
\end{eqnarray}
\item 4-fermion operators\\
These operators simply introduce new vertices involving
four fermions. 
\item Operators that modify triple gauge boson couplings\\
Obviously, $O_W$ is such an operator. The only other one is
$O_{WB}$, which contains a term (see Eq.~(\ref{eq:owb_exp}))  
\begin{equation}
\mbox-a_{WB}v^2gf^{abc}W^{b\mu}W^{c\nu}\partial_\mu B_\nu.
\end{equation}
These operators alter the cross-sections for $e^+e^-\rightarrow
W^+W^-$ which will be dicussed later. 
\item $O_{eB}$ and $O_{eW}$ are discussed later.
\end{enumerate}
\subsection{Direct and indirect corrections}
Now we consider the corrections to the EWPOs. The corrections can
be divided to two categories, direct and indirect. To understand the
two different kinds of
corrections, let us write Eq.~(\ref{eq:Xth0}) in a different way:
\begin{equation}
X_{th}=X_{SM}(g,g',v)+\sum a_i X'_i.\label{eq:Xth1}
\end{equation}
Here $X_{SM}(g,g',v)$ is the SM prediction calculated from the
Lagrangian parameters $g$, $g'$ and $v$. At loop levels, it
also depends on a few other parameters, $m_t$, $m_h$ and $\alpha_s$ which we
take as fixed.

$X'_i$ is the ``direct'' correction from new Feynman diagrams
induced by the operator $O_i$. For example, the four fermion operator
$(\overline e\gamma^\mu e)( \overline q\gamma_\mu q)$ introduces a
4-fermion vertex and therefore a diagram for
 $e^+e^-\rightarrow \overline q q$. A correction to the
$Z$-boson coupling can also be viewed as a new diagram with a new
coupling. 

Less obvious are the ``indirect'' corrections from the shifts to the
input parameters: the SM parameters $g$, $g'$ and $v$ and therefore
$X_{SM}(g,g',v)$ are not directly
accessible. Instead, we derive them from the most precisely measured
observables ($\alpha$, $M_Z$, $G_F$). When calculating the SM predictions
$X_{SM}(\alpha,M_Z,G_F)$, the SM relations between ($g$, $g'$,
$v$), and ($\alpha$, $M_Z$, $G_F$) are
assumed. These relations are altered when the new operators are
included. For example, the operator $O_h$ contributes to the physical
$Z$ mass. Taking into consideration the shifts, we can express ($g$,
$g'$, $v$) as functions of both ($\alpha$, $M_Z$, $G_F$), and the
operator coefficients
$a_i$'s. Since $a_i$'s are small, we use the linear approximation to obtain 
\begin{equation}
X_{SM}[g(M_Z,a_i)]=X_{SM}[g(M_Z)]+\frac{\delta X_{SM}}{\delta a_i}a_i=
X_{SM}[g(M_Z)]+ \frac{\delta X_{SM}}{\delta g}\frac{\delta
  g}{\delta a_i} a_i.
\end{equation}
For illustration, I have simplified the formula by keeping only one
parameter and one operator coefficient. Generalizing to three
parameters and all operator coefficients are straightforward. Note
that $X_{SM}[g(M_Z)]$ is the SM prediction given in the literature,
including loop corrections. The quantity $\frac{\delta X_{SM}}{\delta
  g}\frac{\delta g}{\delta a_i} a_i$ is the ``indirect'' corrections
to the SM  prediction
due to the presence of the new operator $O_i$. When
calculating the indirect corrections, tree level results for $\frac{\delta
  X_{SM}}{\delta g}$ and $\frac{\delta g}{\delta a_i}$ are sufficient,
which simplify our calculations. Combining the indirect corrections
with the direct correction $X'_i$, we arrive at Eq.~(\ref{eq:Xth0})
given in the introduction.

\subsection{Corrections to EWPOs}
We start our calculation from indirect corrections, by establishing
the relations between the parameters in the Lagrangian ($g$, $g'$,
$v$) and the input parameters  ($\alpha$, $M_Z$, $G_F$).
For convenience, we define
\begin{equation}
M_{Z0}=\frac{\sqrt{g^2+g'^2}v}{2},\quad G_{F0}=\frac1{\sqrt{2}v^2},
\quad \alpha_0=\frac{e^2}{4\pi}.
\end{equation}
Gathering the corrections to $\alpha$ and $M_Z$ from 
Eq.~(\ref{eq:piprime}) and (\ref{eq:oh_exp}), we have
\begin{eqnarray}
\alpha& =&\frac{e^2}{4\pi}(1-2v^2s c \, a_{W\!B}),\label{eq:alphashift}\\
 M_Z&=&M_{Z0}(1+v^2sc\, a_{W\!B}+ \frac{v^2 }{4
}a_h).\label{eq:mzshift}
\end{eqnarray}
$G_F$ is obtained from muon lifetime, which is determined by the
effective Lagrangian
\begin{equation}
\mathcal{L}=-2\sqrt{2}G_F\bar
e_L\gamma^\mu\nu_{eL}\bar{\nu}_{\mu L}\gamma_\mu\mu_L.\nonumber
\end{equation}
There are two operators that contribute to the effective Lagrangian, one
from the 4-fermion operator $O_{ll}^t$, and the other one given by the
corrections to $W$-fermion couplings from $O_{hl}^t$. Therefore, 
\begin{equation}
G_F=G_{F0}+\frac{1}{\sqrt{2}}\left( 2 a_{hl}^t -a_{ll}^t\right)\label{eq:gfshift}.
\end{equation}
We can easily solve Eqs.~(\ref{eq:alphashift}), (\ref{eq:mzshift}) and
(\ref{eq:gfshift}) for $g$, $g'$ and $v$. For a given observable, we
 then substitute the solutions in the corresponding SM expression and
expand to linear order in $a_i$'s. For example,
from 
\begin{equation}
M_W=\frac{gv}{2},\label{eq:mwtree}
\end{equation}
we obtain the corrections to $M_W$:
\begin{equation}
\Delta M_W=10^6\left(-1.73a_h-2.08a_{hl}^t+1.04a_{ll}^t-3.80a_{WB}\right),\label{eq:deltamw}
\end{equation}
where $a_i$ is in $\GeV^{-2}$. Then the theoretical prediction for
$M_W$ is
\begin{eqnarray}
M_{W,th}=M_{W,SM}+\Delta M_W=80.376\,\GeV+\Delta M_W.\label{eq:mwth}
\end{eqnarray} 
 Note that the SM prediction $M_{W,SM}$ is the one given in Table
\ref{table:experiments} which contains SM loop corrections. It is {\it not}
calculated with the tree level expression, Eq.~(\ref{eq:mwtree}),
which is only used to calculate  corrections from new physics. 

There is no direct correction to $M_W$, so Eq.~(\ref{eq:mwth})
contains all corrections to $M_W$. The experimental value of $M_W$ is
obtained from both LEP 2\cite{lep2} and Tevatron\cite{WmassTevatron}.

For other observables, we should add the direct corrections as
well. In the following, we will give
both direct corrections and the SM tree level formulae for calculating
indirect corrections (but omit the obvious step of expanding
to linear order in $a_i$).

\begin{enumerate}
\item {\bf$Z$-pole observables}\\ 
The process is $e^+e^-\rightarrow f\overline f$, which was studied around the
$Z$-pole at SLC and LEP 1. SLC also had longitudinal beam polarization.
The measured cross-sections and asymmetries for different center of
mass energies have been translated to pseudo-observables at the
$Z$-pole, which can all be derived from two basic quantities: 
the partial decay width of $Z\rightarrow f\overline f$,  
$\Gamma_{ff}$, and the polarized asymmetry
$A_f$. They are related to the $Z$-fermion couplings $g_V^f$ and
$g_A^f$ defined in eq.~(\ref{eq:gvga}):
\begin{eqnarray}
\Gamma_{ff}&=&\frac{e^2M_Z}{48\pi s^2
c^2}(g_V^{f2}+g_A^{f2}),\\
 A_f&=&2\frac{g_V^f/g_A^f}{1+(g_V^f/g_A^f)}.
\end{eqnarray}
In the above equations, $g_V=g_{V0}+\Delta g_V$ and $g_A=g_{A0}+\Delta
g_A$, where $g_{V0}$ and $g_{A0}$ are given in (\ref{eq:gv0ga0}), and
$\Delta g_V$ and $\Delta g_A$ are given in (\ref{eq:dgvdga}). Note
that $g_V$ and $g_A$ contain both direct and indirect corrections. 
From $\Gamma_{ff}$ and $A_f$, we derive the following observables: 
\begin{enumerate}
\item Total decay width $\Gamma_Z=\sum_f\Gamma_{ff}$.
\item Total hadronic cross-section
\begin{equation}
\sigma_h^0=\frac{12\pi}{M_Z^2}\frac{\Gamma_{ee}\Gamma_{had}}{\Gamma_Z},
\end{equation}
where $\Gamma_{had}$ is the partial decay width to hadrons and
$\Gamma_{ee}$ is the partial decay width to $e^+e^-$.
\item The ratios
\begin{equation}
R^0_e=\Gamma_{had}/\Gamma_{ee},\ \ R^0_\mu=\Gamma_{had}/\Gamma_{\mu\mu}, \ \
R^0_\tau=\Gamma_{had}/\Gamma_{\tau\tau}.
\end{equation}
\item The pole asymmetries $A_{fb}^{0,f}$, ($f=e,\mu,\tau$) related to
  $A_f$ by
\begin{equation}
A_{fb}^{0,f}=\frac34A_eA_f.
\end{equation}
\item The polarized asymmetries $A_e$, $A_\mu$, $A_\tau$ from both
  LEP 1 and SLC.
\item Heavy flavor observables including 
\begin{equation}
R_b^0=\Gamma_{b\bar b}/\Gamma_{had}, R_c^0=\Gamma_{c\bar
  c}/\Gamma_{had}, A_{fb}^{0,b}, A_{fb}^{0,c},A_b,A_c.
\end{equation}
\item Hadronic charge asymmetry $\langle Q_{fb}\rangle$.
The result is translated to the leptonic effective electroweak mixing
angle, 
\begin{equation}
\sin^2\theta^{\mbox{lept}}_{\mbox{eff}}=\frac14\left(1-\frac{g_{V}^e}{g_{A}^e}\right).
\end{equation}
\end{enumerate}
Note that at the $Z$-pole, the denominator of the $Z$ propagator is
dominated by $iM_Z\Gamma_Z$, which is imaginary relative to
the 4-fermion operator contributions. Therefore, 4-fermion operators do
not interfere with the $Z$-pole measurements and there is no
contributions linear in the corresponding $a_i$'s.  
\item {\bf$\nu$-Nucleon scattering}\\
This includes a variety of deep inelastic experiments performed at
CDHS\cite{CDHS}, CHARM\cite{CHARM}, CCFR\cite{CCFR} and
NuTeV\cite{NuTeV}. The experiments measured cross sections for both
the neutral current 
(NC), $\nu_\mu+N\rightarrow \nu_{\mu}+N$, and the charged current
(CC) $\nu_\mu+N\rightarrow \mu+X$. The experimental
results are usually translated to the effective couplings between the $Z$
boson, and the up and down quarks: $g_{L,eff}^u$, $g_{R,eff}^u$,
$g_{L,eff}^d$ and $g_{R,eff}^d$. They are effective in the sense that
they only apply to the $\nu$-nucleon scattering observables. Since
that these experiments were performed at
energies far below the $Z$-pole\footnote{However, note that the
  results have been renormalized to the scale of $M_Z$.}, the effective couplings receive
contributions from 
both 4-fermion operators and operators modifying gauge-fermion
couplings.
In order to cancel out uncertainties from the
strong interaction, what is measured is actually the ratio between the NC
 and the CC cross sections. Therefore, both
corrections to the NC and CC need to be considered.

For the CC, comparing the effective Lagrangian in the SM
\begin{equation}
\mathcal{L}_{CC}^{eff}=-2\sqrt{2}G_F(\overline{\nu}_{\mu L}\gamma^\mu
\mu_L)(d_L\gamma_\mu u_L),
\end{equation}
and the one with new operators
\begin{equation}
\mathcal{L}_{CC}^{eff}=\left[-2\sqrt{2}G_{F0}(1+a_{hl}^{t}v^2+a_{hq}^{t})+
 2a_{lq}^{t}\right] (\overline{\nu}_{\mu L}\gamma^\mu \mu_L)(d_L\gamma_\mu u_L),
\end{equation}
we see that it is modified by an overall factor 
\begin{equation}
F_{cc}=\frac{G_{F0}}{G_F}\left[(1+a_{hl}^{t}v^2+a_{hq}^{t}v^2)-\frac{a_{lq}^{t}}{2\sqrt{2}G_F}\right].
\end{equation}

The effective Lagrangian for the NC is
\begin{eqnarray}
\mathcal{L}_{NC}^{eff}&=&-\frac{e^2}{2s^2c^2 M_Z^2}\overline\nu_\mu\gamma^\mu(g_V^\nu-g_A^\nu\gamma^5)\nu_\mu\sum_q(g_L^q\bar q_L\gamma_\mu q_L+
   g_R^q\bar q_R\gamma_\mu q_R\nonumber)\\
&&+a_{lq}^s(\bar \nu_{\mu L}\gamma^\mu\nu_{\mu L})(\bar u_L\gamma_\mu u_L+\bar d_L\gamma_\mu d_L)\nonumber\\
&&+a_{lq}^t(\bar \nu_{\mu L}\gamma^\mu\nu_{\mu L})(\bar u_L\gamma_\mu u_L-\bar d_L\gamma_\mu d_L)\nonumber\\
&&+a_{lu}(\bar \nu_{\mu L}\gamma^\mu\nu_{\mu L})(\bar u_R\gamma_\mu u_R)\nonumber\\
&&+a_{ld}(\bar \nu_{\mu L}\gamma^\mu\nu_{\mu L})(\bar d_R\gamma_\mu
   d_R),
\end{eqnarray}
where 
\begin{equation}
g_L^q=\frac{g^q_V+g^q_A}{2},\quad g_R^q=\frac{g^q_V-g^q_A}{2}.
\end{equation}
Note that $g_V^\nu=g_A^\nu$ holds for all new operator contributions. 
Therefore, all new physics contributions can be absorbed in the
effective couplings $g^{u,d}_{L,eff}$ and $g^{u,d}_{R,eff}$,
\begin{equation}
\mathcal{L}_{NC,eff}=-2\sqrt{2}G_F\overline{\nu}_{\mu
  L}\gamma^\mu\nu_{\mu L}\sum_{q=u,d}(g^{q}_{L,eff}
\overline q_L\gamma_\mu q_L+g^{q}_{R,eff}\overline q_R\gamma_\mu q_R).
\end{equation}
with
\begin{eqnarray}
g_{L,eff}^u&=&\frac1{F_{CC}}\frac{G_{F0}}{G_F}\frac{M_{Z0}^2}{M_{Z}^2}
\left[2g_V^{\nu_\mu}\left(g_L^u-\frac{a_{lq}^{s}}{2\sqrt{2}G_F}-
\frac{a_{lq}^{t}}{2\sqrt{2}G_F}\right)\right],\nonumber\\
g_{L,eff}^d&=&\frac1{F_{CC}}\frac{G_{F0}}{G_F}\frac{M_{Z0}^2}{M_{Z}^2}
\left[2g_V^{\nu_\mu}\left(g_L^d-\frac{a_{lq}^{s}}{2\sqrt{2}G_F}+
\frac{a_{lq}^{t}}{2\sqrt{2}G_F}\right)\right],\nonumber\\
g_{R,eff}^u&=&\frac1{F_{CC}}\frac{G_{F0}}{G_F}\frac{M_{Z0}^2}{M_{Z}^2}
\left[2g_V^{\nu_\mu}\left(g_R^u-\frac{a_{lu}}{2\sqrt{2}G_F}\right)\right],\nonumber\\
g_{R,eff}^d&=&\frac1{F_{CC}}\frac{G_{F0}}{G_F}\frac{M_{Z0}^2}{M_{Z}^2}
\left[2g_V^{\nu_\mu}\left(g_R^d-\frac{a_{ld}}{2\sqrt{2}G_F}\right)\right].   
\end{eqnarray}
The measured quantities are expressed in terms of the effective
couplings:
\begin{equation}
g_{L}^2=g_{L,eff}^{u2}+g_{L,eff}^{d2},\quad
g_{R}^2=g_{R,eff}^{u2}+g_{R,eff}^{d2}.
\end{equation} 
\item {\bf$\nu$-$e$ scattering}\\
$\nu-e$ scattering was performed at CHARM II\cite{CHARMII}. Similarly,
  the corrections are obtained by considering the effective Lagrangian:
\begin{equation}
\mathcal{L}_{\nu
  e,eff}=-\sqrt{2}G_F\overline{\nu}_L\gamma^\mu\nu_L\overline
e\gamma_\mu(g_{V,eff}^{\nu e}-g_{A,eff}^{\nu e}\gamma^5)e,
\end{equation}
where $g^{\nu e}_{V,eff}$ and $g^{\nu e}_{A,eff}$ are measured in the
experiment and related to our operators as
\begin{eqnarray}
g_{V,eff}^{\nu e}&=&\frac{M_{Z0}^2}{M_Z^2}\frac{G_{F0}}{G_F}2g_V^\nu\left(g_V^e-\frac{a_{ll}^s}{2\sqrt{2}G_F}+\frac{a_{ll}^t}{2\sqrt{2}G_F}
   -\frac{a_{le}}{2\sqrt{2}G_F}\right),\nonumber\\
g_{A,eff}^{\nu e}&=&\frac{M_{Z0}^2}{M_Z^2}\frac{G_{F0}}{G_F}2g_V^\nu\left(g_A^e-\frac{a_{ll}^s}{2\sqrt{2}G_F}+\frac{a_{ll}^t}{2\sqrt{2}G_F}
   +\frac{a_{le}}{2\sqrt{2}G_F}\right).
\end{eqnarray}
\item {\bf Weak charge}\\
The weak charges for Cs and Tl are measured in the atomic parity
violation experiments\cite{QWCs}\,\cite{QWTl}. The effective Lagrangian
is conventionally written in terms of 
$C_{1q}$ and $C_{2q}$:
\begin{equation}
\mathcal{L}=\frac{G_F}{\sqrt{2}}\sum_q\left[C_{1q}(\bar e\gamma^\mu\gamma^5 e)(\bar q\gamma_\mu q)+C_{2q}(\bar e\gamma^\mu e)(\bar q\gamma_\mu\gamma^5q)\right].
\end{equation}
We only need the corrections to $C_{1u}$ and $C_{1d}$:
\begin{eqnarray}
C_{1u}&=&\frac{G_{F0}}{G_F}\frac{M_{Z0}^2}{M_Z^2}2g_A^eg_V^u+\frac1{2\sqrt{2}G_F}
(-a_{lq}^s+a_{lq}^t+a_{eu}+a_{eq}-a_{lu}),\nonumber\\
C_{1d}&=&\frac{G_{F0}}{G_F}\frac{M_{Z0}^2}{M_Z^2}2g_A^eg_V^d+\frac1{2\sqrt{2}G_F}
(-a_{lq}^s-a_{lq}^t+a_{ed}+a_{eq}-a_{ld}).
\end{eqnarray}
Then the weak charge is given by
\begin{equation}
Q_W(Z,N)=-2[(2Z+N)C_{1u}+(Z+2N)C_{1d}],
\end{equation}
where $Z$ and $N$ are respectively proton number and neutron number of the atom. 
\item {\bf$e^+e^-\rightarrow f\overline f$ at LEP 2}\\
The observables are differential cross-sections for
fermion pair production. In the following, we will give the formulae for the
matrix elements from which the cross-sections can be calculated. Unlike LEP
1, LEP 2 experiments were performed above the $Z$-pole, therefore,
4-fermion operators also contribute.
\begin{enumerate}
\item $e^+e^-\rightarrow f\overline f (f\neq e)$\cite{lep2}\\
\begin{eqnarray}
\mathcal{M}&=&\mathcal{M}_Z+\mathcal{M}_\gamma+\mathcal{M}_{4-fermi}\nonumber\\
&=&\frac{e^2}{(p+p')^2-M_Z^2+i\Gamma_ZM_Z}[\frac{1}{4c^2s^2}\bar
v(p')\gamma^\mu(g_V^e-\gamma^5g_A^e)u(p)\bar
u(k)\gamma_\mu(g_V^f-\gamma^5g_A^f)v(k')]\nonumber\\
&&+\frac{-e^2Q}{(p+p')^2}[\bar v(p')\gamma^\mu u(p)][\bar
u(k)\gamma_\mu v(k')]\nonumber\\
&&+a_1[\bar v(p')\gamma^\mu u(p)][\bar u(k)\gamma_\mu v(k')]\nonumber\\
&&+a_2[\bar v(p')\gamma^\mu \gamma^5u(p)][\bar u(k)\gamma_\mu v(k')]\nonumber\\
&&+a_3[\bar v(p')\gamma^\mu u(p)][\bar u(k)\gamma_\mu\gamma^5 v(k')]\nonumber\\
&&+a_4[\bar v(p')\gamma^\mu\gamma^5 u(p)][\bar
u(k)\gamma_\mu\gamma^5 v(k')],
\end{eqnarray}
where $p,p',k,k'$ are the momenta of the incoming $e^+e^-$ and the
outgoing fermions. $\mathcal{M}_Z$ and $\mathcal{M}_\gamma$ are
$s$-channel contributions from $Z$ and photon
exchange. $\mathcal{M}_{4-fermi}$ denotes contributions from 4-fermion
operators, where depending on the fermion $f$,
the couplings $a_1,a_2,a_3,a_4$ are related to the operator
coefficients by
\begin{eqnarray}
a1_\mu&=& 1/4(a_{ll}^s - a_{ll}^t + 2a_{le} + a_{ee}), \nonumber\\
a2_\mu&=& 1/4(-a_{ll}^s +a_{ll}^t + a_{ee}),\nonumber\\
a3_\mu &=& 1/4(-a_{ll}^s + a_{ll}^t + a_{ee}),\nonumber\\
a4_\mu &=& 1/4(a_{ll}^s - a_{ll}^t - 2a_{le} + a_{ee});
\end{eqnarray}
\begin{eqnarray}
a1_u& =& 1/4(a_{lq}^s - a_{lq}^t + a_{lu} + a_{eq} + a_{eu}),\nonumber\\
 a2_u& =&1/4(-a_{lq}^s + a_{lq}^t - a_{lu} + a_{eq} + a_{eu}),\nonumber\\
 a3_u& =& 1/4(-a_{lq}^s + a_{lq}^t + a_{lu} -a_{eq} + a_{eu}),\nonumber\\
 a4_u& =& 1/4(a_{lq}^s - a_{lq}^t - a_{lu} - a_{eq} + a_{eu});
\end{eqnarray}
\begin{eqnarray}
a1_d& =& 1/4(a_{lq}^s + a_{lq}^t + a_{ld} + a_{eq} + a_{ed}),\nonumber\\
 a2_d& =&1/4(-a_{lq}^s - a_{lq}^t - a_{ld} + a_{eq} + a_{ed}),\nonumber\\
 a3_d& =& 1/4(-a_{lq}^s - a_{lq}^t + a_{ld} -
a_{eq} + a_{ed}),\nonumber\\
a4_d &= &1/4(a_{lq}^s + a_{lq}^t - a_{ld} - a_{eq} + a_{ed}).
\end{eqnarray}
\item $e^+e^-\rightarrow e^+e^-$\cite{OPALfpair}\\
For $f=e$, there are extra contributions from $t$-channel
diagrams. The matrix element is
\begin{eqnarray}
\mathcal{M}&=&\mathcal{M}_\gamma+\mathcal{M}_{Z}+
\mathcal{M}_{4-fermi} \nonumber\\
&=&\frac{e^2}{(p-k)^2}\bar u(k)\gamma^\mu u(p)\bar
v(p')\gamma_\mu v(k')\nonumber\\
&&-\frac{e^2}{(p+p')^2}\bar v(p')\gamma^\mu u(p)\bar
u(k)\gamma_\mu v(k')\nonumber\\
&&+\frac{e^2}{(p-k)^2-M_Z^2+i\Gamma_ZM_Z}\frac1{4c^2s^2}\bar
u(k)\gamma^\mu(g_V^e-\gamma^5 g_A^e)u(p)\bar
v(p')\gamma_\mu(g_V^e-\gamma^5g_A^e)v(k')\nonumber\\
&&-\frac{e^2}{(p+p')^2-M_Z^2+i\Gamma_ZM_Z}\frac1{4c^2s^2}\bar
v(p')\gamma^\mu(g_V^e-\gamma^5 g_A^e)u(p)\bar
u(k)\gamma_\mu(g_V^e-\gamma^5g_A^e)v(k')\nonumber\\
&&+a_1[ \bar u(k)\gamma^\mu u(p)\bar v(p')\gamma_\mu v(k')-\bar
v(p')\gamma^\mu u(p)\bar u(k)\gamma_\mu v(k')]\nonumber\\
&&+a_2[ \bar u(k)\gamma^\mu \gamma^5u(p)\bar v(p')\gamma_\mu
v(k')-\bar
v(p')\gamma^\mu \gamma^5u(p)\bar u(k)\gamma_\mu v(k')]\nonumber\\
&&+a_3[ \bar u(k)\gamma^\mu u(p)\bar v(p')\gamma_\mu\gamma^5
v(k')-\bar
v(p')\gamma^\mu u(p)\bar u(k)\gamma_\mu\gamma^5 v(k')]\nonumber\\
&&+a_4[ \bar u(k)\gamma^\mu\gamma^5 u(p)\bar v(p')\gamma_\mu
\gamma^5v(k')-\bar v(p')\gamma^\mu\gamma^5 u(p)\bar
u(k)\gamma_\mu\gamma^5 v(k')],
\end{eqnarray} 
where
\begin{eqnarray}
a1&= &1/4(a_{ll}^s + a_{ll}^t + 2a_{le}+a_{ee})\nonumber\\
a2& =& 1/4(-a_{ll}^s - a_{ll}^t +a_{ee})\nonumber\\
a3& =& 1/4(-a_{ll}^s - a_{ll}^t +a_{ee})\nonumber\\
a4& =& 1/4(a_{ll}^s + a_{ll}^t - 2a_{le}+a_{ee}).
\end{eqnarray}
\end{enumerate}
\item {\bf $e^+e^-\rightarrow W^+W^-$ at LEP 2\cite{L3Wpair}}\\
There are two diagrams contributing to this process,
one with $t$-channel neutrino exchange and the other one with
$s$-channel $\gamma$ or $Z$ exchange. The former involves $W$-fermion
couplings and therefore receives corrections from (\ref{eq:wff}). The
$\gamma$ or $Z$ exchange diagram involves triple gauge boson
couplings, which are modified by the operators $O_{WB}$ and $O_W$. 
The relevant couplings are given by
 \begin{equation}
 \label{eq:3gauge}
\Delta \mathcal{L}= ia_{W\!B} v^2 gW^+_\mu W^-_\nu(c A^{\mu\nu}-s
  Z^{\mu\nu})+6ia_WW^{-\mu}_\nu
  W^{+\lambda}_\mu(sA^\nu_\lambda+cZ_\lambda^\nu).
 \end{equation}
The tree-level differential cross-section for $e^+e^-\rightarrow
W^+W^-$ is calculated in Ref.~\refcite{triplegauge} for arbitrary triple gauge
boson couplings, which are parametrized as 
\begin{eqnarray}
\frac{\mathcal{L}_{WWV}}{g_{WWV}}&=&ig_1^V(W_{\mu\nu}^+W^{-\mu}V^\nu-
W_{\mu\nu}^-W^{+\mu}V^\nu)
+i\kappa_VW^+_\mu W_\nu^-V^{\mu\nu}+\frac{i\lambda_V}{\Lambda_\chi^2}
W_{\mu\nu}^+W^{-\nu}_\rho
V^{\rho\mu}\nonumber\\
&&-g_4^VW_\mu^+W_\nu^-(\partial^\mu V^\nu+\partial^\nu V^\mu)
+g_5^V\epsilon^{\mu\nu\rho\lambda}\left[W_\mu^+(\partial_\rho
  W_\nu^-)-(\partial_\rho W_\mu^+)W_\nu^-\right]V_\lambda\nonumber\\
&& +i\tilde K_VW_\mu^+W_\nu^-\tilde
V^{\mu\nu}+\frac{i\tilde\lambda_V}{\Lambda_\chi^2}
W^+_{\mu\nu}W^{-\nu}_\rho\tilde V^{\rho\mu},\label{eq:triple}
\end{eqnarray}
where $V=\gamma,Z$ and $g_{WW\gamma}=-e$, $g_{WWZ}=-ec/s$.
Our effective couplings, Eq.~(\ref{eq:3gauge}),
correspond to the terms multiplying
 $\kappa_V$ and $\lambda_V$. To obtain the cross-section we substitute
\begin{eqnarray}
 \Delta\kappa_\gamma&=&\frac{ v^2 c}{s} \,a_{W\!B} ,\nonumber\\
 \Delta\kappa_Z&=&-\frac{v^2s}{c} \, a_{W\!B},\nonumber\\
 \Delta\lambda_\gamma&=&\Delta\lambda_Z=\frac{3v^2 g}{2} \,a_W .
\end{eqnarray}
where $\Delta$'s denote the deviations from the SM values.  
\item {\bf Muon $g-2$\cite{gminus2}}\\
 As mentioned, the operators
$O_{eB}$ and $O_{eW}$ contribute to
$g_\mu-2$ at tree level,
\begin{equation}
a_\mu=\frac{2\sqrt{2}vm_\mu}{g}a_{eW}-\frac{2\sqrt{2}vm_\mu}{g'}a_{eB}.\label{eq:g-2}
\end{equation}
Due to the high precision of this
measurement, these corrections certainly cannot be introduced at tree
level for TeV scale physics. Therefore, one should 
assume $a_{eB}$ and $a_{eW}$ vanish and consider loop corrections from
the other operators. These include corrections to the SM loop diagram
from shift of input parameters, and corrections from new loop diagrams
induced by the effective operators. This has been done in
Ref.~\refcite{g-2}. In this
review, we will simply assume that all corrections have been absorbed into
$a_{eB}$ and $a_{eW}$ and put constraints on them using Eq.~(\ref{eq:g-2}).
\end{enumerate}
\section{Constraints}
\label{sec:constraints}
The constraint on each individual operator is given in
Table \ref{table:lambdas}. The bounds on $O_{eB}$ and $O_{eW}$ is
calculated from muon $g-2$ with experimental values given in Table
\ref{table:experiments}. Bounds on the other operators are taken from
Ref.~\refcite{Han+Skiba}. 
\begin{table}[h]
\tbl{Bounds on operator coefficients at 90\% CL. The operator
  coefficient $a_i$ is bounded by the interval $[-1/\Lambda_{min}^2,
    1/\Lambda_{max}^2]$, except for $a_{eB}$ and
    $a_{eW}$ which are multiplied by an extra factor of
    $g'm_\mu/16\pi^2 v$ and $gm_\mu/16\pi^2 v$
    respectively. The $\Lambda_i$ (in TeV) shown in the table is the average of
    $\Lambda_{min}$ and $\Lambda_{max}$. When calculating each
  $\Lambda_i$, all other operator coefficients are set to zero. }
{\begin{tabular}{cccccccc}
\hline 
$O_i$&$\Lambda_i$&$O_i$&$\Lambda_i$&$O_i$&$\Lambda_i$&$O_i$&$\Lambda_i$\\
\hline
$O_{WB}$&12.6&$O_h$&6.8&$O_{ll}^s$&3.6&$O_{ll}^t$&8.8\\
$O_{lq}^s$&5.4&$O_{lq}^t$&6.2&$O_{le}$&4.3&$O_{qe}$&5.5\\
$O_{lu}$&3.8&$O_{ld}$&4.0&$O_{ee}$&3.5&$O_{eu}^s$&4.3\\
$O_{ed}$&3.9&$O_{hl}^s$&11.6&$O_{hl}^t$&11.5&$O_{hq}^s$&6.4\\
$O_{hq}^t$&9.2&$O_{hu}$&4.5&$O_{hd}$&4.3&$O_{he}$&9.7\\
$O_{W}$&0.9&$O_{eB}$&0.4&$O_{eW}$&0.4\\
\hline
\end{tabular}\label{table:lambdas}}
\end{table}

From Table \ref{table:lambdas} we see that many of the operators are
tightly constrained. At 90\% CL, there are still quite a few bounds around
10 TeV, which manifest the ``LEP paradox''\cite{paradox}. 

The operator $O_W$ is not constrained as well as others, due to the
low statistics in the measurement of $e^+e^-\rightarrow W^+W^-$ cross sections. 
Therefore, in the literature, this measurement is often neglected from
the list of EWPOs. Nevertheless, it gives the unique constraint of
$O(\TeV)$ if the new physics only modifies the triple gauge couplings.

We also see from the bounds on $a_{eB}$ and $a_{eW}$ that, the muon
$g-2$ measurement does not give us stringent constraints if
the new physics contribution is suppressed by a loop factor and the
muon mass. See Ref.~\refcite{g-2} for exceptions, as well as an analysis
for the non-linear case.

Although it is clear from Table \ref{table:lambdas} how well each of the
operators is constrained, in practice, constraints on individual operators
are not useful. The reason is that we usually obtain multiple
operators when the new physical states are integrated out. Their contributions
to the observables are correlated. Therefore, in order to obtain the
bounds on a new model, the full $\chi^2$ distribution should be
used. Including experimental correlations, Eq.~(\ref{eq:chi2uncorr})
is modified to 
\begin{equation}
 \label{eq:chi2corr}
  \chi^2(a_i)=\sum_{p,q}(X_{th}^p(a_i)-X_{exp}^p)(\sigma^2)^{-1}_{pq}
  (X_{th}^q(a_i)-X_{exp}^q),
\end{equation}
where $\sigma^2$ is the error matrix, defined from the standard
deviations $\sigma_p$ and the correlation matrix $\rho_{pq}$ as
\begin{equation}
 \sigma^2_{pq}=\sigma_p\rho_{pq}\sigma_q.
\end{equation} 
The numerical result of the $\chi^2$ distribution, for operators
(\ref{eq:owb})-(\ref{eq:ow}), is given in Ref.~\refcite{Han+Skiba}.
After obtaining the $\chi^2$, it is straightforward to calculate
constraints for a given model once the heavy particles are integrated
out and the coefficients $a_i$'s are obtained. See
Ref.~\refcite{lhbounds}-\refcite{chbounds} for a few examples on the little
Higgs models\cite{lhreview} and models with a warped extra-dimension.
\section{Discussions}
\label{sec:discussions}
\subsection{Oblique corrections}
The oblique corrections are modifications to the gauge boson
propagators. We discuss  in this section the relations between the oblique
parameters and the effective operators.
\subsubsection{$S$, $T$ and $U$}
The well known $S$, $T$ and $U$ parameters\cite{STU} are defined by
\begin{eqnarray}
\frac{\alpha}{4s^2c^2}S&=&\Pi'_{ZZ}-\frac{c^2-s^2}{cs}\Pi'_{Z\gamma}-
\Pi'_{\gamma\gamma},\nonumber\\ 
\alpha T&=&\frac{\Pi_{WW}(0)}{M_W^2}-\frac{\Pi_{ZZ}(0)}{M_Z^2},\nonumber\\
\frac{\alpha}{4s^2}(S+U)&=&\Pi'_{WW}-\frac{c}{s}\Pi'_{Z\gamma}-
\Pi'_{\gamma\gamma}. \label{eq:stu}
 \end{eqnarray}
where the corrections only include new physics contributions.
In our framework, the operators $O_h$ and $O_{WB}$ modify gauge
boson propagators. Comparing with Eqs.~(\ref{eq:piprime}) and
(\ref{eq:pizz0}), we have
\begin{equation}
S=\frac{4scv^2a_{W\!B}}{\alpha},\quad T=-\frac{v^2}{2\alpha} a_h,
\quad U=0.
\end{equation}
 We
see that except for some constant factors, $S$ and $T$ are equivalent
to $a_{WB}$ and $a_h$. Moreover, there is no dimension-6 operator
corresponding to the $U$ parameter. It turns out that the
dimension-8 operator $(h^\dag W^{a\mu\nu}h)( h^\dag W_{\mu\nu}^ah)$
corresponds to the $U$ parameter. By our power counting, its
contribution is negligible. This explains why $U$ parameter is usually
not significant in weakly coupled extensions of the SM. Nevertheless,
as we will see, in the non-linearly realized electroweak symmetry
case, $U$ can be important due to a
different power counting. Therefore, we discuss below the
corrections from the $U$ parameter to the EWPOs. 

From Eqs.~(\ref{eq:stu}), we have
\begin{equation}
\frac{\alpha}{4s^2}U=\Pi'_{WW}-\Pi'_{W^3W^3}.
\end{equation}
When $\Pi'_{WW}=\Pi'_{W^3W^3}$, which corresponds to multiplying the
kinetic term $-W^{\mu\nu}W_{\mu\nu}/4$ in the SM Lagrangian by an
overall factor, there is
no visible correction. Therefore, we can always trade $\Pi'_{W^3W^3}$
for $\Pi'_{WW}$, which can be absorbed into field redefinition of $W^+$
and $W^-$:
\begin{equation}
W^+\rightarrow W^+(1-\Pi'_{WW})^{\frac12}.
\end{equation}
The field redefinition affects the $W$ mass,
\begin{equation}
\Delta M_W^2=M_W^2 \Pi'_{WW}=M_W^2\frac{\alpha U}{4s^2},
\end{equation}
and $W$-fermion couplings
\begin{equation}
g\rightarrow g(1+\frac{\alpha U}{8s^2}).
\end{equation}
Note that $U$ does not affect low energy experiments such as muon
lifetime ($G_F$) or
charged current in DIS experiments because the matrix element is
proportional to $g^2/M_W^2$, where $U$ cancels out.

As shown in Ref.~\refcite{pdg}, the $S$, $T$ and $U$ parameters are
related to other parameters $(S_i, hi, \hat\epsilon_i)$\cite{Si}\cdash\cite{epsiloni} as
\begin{eqnarray}
T&=&h_V=\hat\epsilon_1/\alpha,\nonumber\\
S&=&h_{AV}=S_Z=4s^2\hat\epsilon_3/\alpha,\nonumber\\
U&=&h_{AW}-h_{AZ}=S_W-S_Z=-4s^2\hat\epsilon_2/\alpha,
\end{eqnarray}
which can be used to relate our operators to these parameters as well.

\subsubsection{$W$ and $Y$}
The list of oblique parameters is extended in Ref.~\refcite{WY} to
include two more important parameters: $W$ and $Y$. In the effective
theory language, these parameters correspond to the dimension-6
operators 
\begin{equation}
O_{BB}=\frac12(\partial_\rho B_{\mu\nu})^2, \ \ O_{WW}=\frac12(D_\rho
W_{\mu\nu}^a)^2.
\end{equation}
By Bianchi identities such as $\partial_\rho  B_{\mu\nu}+\partial_\mu
B_{\nu\rho}+\partial_\nu B_{\rho\mu}=0$, they are equal to
$(\partial_\nu B^{\nu\mu})^2$ and $(D_\nu W^{\nu\mu a})^2$
respectively\cite{disguise}.
These operators are not independent from other operators defined
in Ref.~\refcite{Buchmuller+Wyler}. They can be related by the
equations of motion.  Squaring Eqs.~(\ref{eq:weom}) and
(\ref{eq:beom}), we see that each of $O_{WW}$ and
$O_{BB}$ is equivalent to a combination of operators already included in out
list.  

\subsubsection{Disguise the oblique corrections}
Another interesting consequence of the equations of motion is that,
they can be used to ``disguise'' the oblique corrections\cite{disguise},
that is, transfer oblique corrections to
equivalent non-oblique corrections. Multiplying Eq.~(\ref{eq:weom}) by
$(ih^\dag\sigma^aD^\nu h+h.c.)$ and Eq.~(\ref{eq:beom}) by $(ih^\dag
D^\nu h+h.c.)$, we have 
\begin{eqnarray}
 &&-\frac{g}{2}O_{WB} +2g'O_{h} + g'  O_{hf}^Y
         = 2 i B_{\mu \nu} D^\mu h^\dagger D^\nu h, 
 \nonumber\\
 &&  - g'O_{WB}  +g(O_{hl}^t+O_{hq}^t)
         = 4 i W^a_{\mu \nu} D^\mu h^\dagger \sigma^a D^\nu h, \label{eq:disguise}
\end{eqnarray}
where we have neglected operators that do not contribute to EWPOs, and
$O_{hf}^Y=\sum_f Y_f O_{hf}^s =\frac{1}{6} O_{hq}^s-\frac{1}{2}
O_{hl}^s + \frac{2}{3} O_{hu} -\frac{1}{3} O_{hd} - O_{he}$.
The left-hand side of Eqs.~(\ref{eq:disguise}) is the two oblique
operators corresponding to $S$ and $T$, plus operators that modify
gauge-fermion operators. The right-hand side is two operators that
modify triple gauge boson couplings, which is only constrained by
$e^+e^-\rightarrow W^+W^-$ cross sections. Therefore, the combinations
of operators on the left-hand side of Eqs.~(\ref{eq:disguise}) are
relatively weakly constrained. If $e^+e^-\rightarrow W^+W^-$ data is
neglected, there are no constraints at all. Similar observations for
the  non-linear
case is made in Ref.~\refcite{disguise}.

The merit of the oblique parameters is that they are all stringently
constrained by the experiments. When the new physics corrections are
all oblique or the non-oblique corrections are sub-dominant, it is
sufficient to consider oblique parameters only. Nevertheless, we
see from Table \ref{table:lambdas} that some non-oblique operators are
constrained as tightly as the oblique parameters, which are indispensable
if appearing in the effective Lagrangian. In this case, an analysis using the
full set of operators is required. Alternatively, one may use a
subset which contains the most constrained operators (or operator
combinations) and still get sensible constraints\cite{minimal}.  
\subsection{Relax the flavor symmetry}
So far, we have always assumed flavor universality for the effective
operators. Although it is convenient and arises in many models beyond
the SM, it is not necessary. Bounds on flavor violation for the first two
generations are very stringent, therefore it is hard to break
universality for the first two generations without introducing
conflicts with the experiments. However, it is possible to break
flavor universality for the third generation.  This is
due to two reasons. First, the precisions of experiments involving the
third generation is not as good as the first two generations. Second,
in the CKM matrix, the (13) and (23) components are much smaller than
the (12) component. Therefore the gauge eigenstate is approximately
aligned with the flavor eigenstate in the third generation.

Given the above observation, it is possible that the new physics couples
to the third generation differently from the first two
generations, see Ref.~\refcite{simplest} and \refcite{gaugeext} for a
few examples. Accordingly, the flavor
symmetry must be relaxed, which alters the counting of the effective
operators. We remind the reader that in  Ref.~\refcite{Han+Skiba},
flavor universality is guaranteed by imposing a $U(3)^5$ symmetry. This
flavor symmetry is relaxed in Ref.~\refcite{Han} to $[U(2)\times
  U(1)]^5$, with the $U(2)$'s acting on the
first two generations and $U(1)$'s the third. Correspondingly, the
operators in (\ref{eq:owb})-(\ref{eq:ow}) are still present, but the
fermions $q$, $l$, $u$, $d$ and $e$ are contracted over the first two
generations only. In addition, there are new operators
involving only the third generation. Denoting the third generation
fermions by $Q$, $L$, $b$ and $\tau$, these are
\begin{enumerate}
\item{Four-fermion operators:}
\begin{eqnarray}
  &&
  O_{lL}^s= (\overline{l} \gamma^\mu l) (\overline{L} \gamma_\mu L), \ \ \
  O_{lL}^t= (\overline{l} \gamma^\mu \sigma^a l) (\overline{L} \gamma_\mu \sigma^a L),
  \nonumber\\ &&
  O_{lQ}^s= (\overline{l} \gamma^\mu l) (\overline{Q} \gamma_\mu Q), \ \ \
  O_{lQ}^t= (\overline{l} \gamma^\mu \sigma^a l) (\overline{Q} \gamma_\mu \sigma^a Q),
  \nonumber\\ &&
  O_{Le}= (\overline{L} \gamma^\mu L) (\overline{e} \gamma_\mu e),  \ \ \
  O_{l\tau}= (\overline{l} \gamma^\mu l) (\overline{\tau} \gamma_\mu \tau),
  \nonumber\\ &&
  O_{Qe}=(\overline{Q} \gamma^\mu Q) (\overline{e} \gamma_\mu e),\ \ \
  O_{lb}= (\overline{l} \gamma^\mu l) (\overline{b} \gamma_\mu b),
  \nonumber\\ &&
  O_{e\tau}=(\overline{e} \gamma^\mu e) (\overline{\tau} \gamma_\mu \tau), \ \ \
  O_{eb}=(\overline{e} \gamma^\mu e) (\overline{b} \gamma_\mu b);
  \label{op:4f2}
\end{eqnarray}
\item{Operators modifying gauge-fermion couplings:}
\begin{eqnarray}
  &&
  O_{hL}^s = i (h^\dagger D^\mu h)(\overline{L} \gamma_\mu L) + {\rm h.c.}, \ \ \
  O_{hL}^t = i (h^\dagger \sigma^a D^\mu h)(\overline{L} \gamma_\mu \sigma^a L)+ {\rm h.c.},
  \nonumber\\ &&
  O_{hQ}^s = i (h^\dagger D^\mu h)(\overline{Q} \gamma_\mu Q)+ {\rm h.c.}, \ \ \
  O_{hQ}^t = i (h^\dagger \sigma^a D^\mu h)(\overline{Q} \gamma_\mu \sigma^a Q)+ {\rm h.c.},
  \nonumber\\ &&
  O_{h\tau} = i (h^\dagger D^\mu h)(\overline{\tau} \gamma_\mu \tau)+ {\rm h.c.},\ \ \
  O_{hb} = i (h^\dagger D^\mu h)(\overline{b} \gamma_\mu b)+ {\rm h.c.}.
  \label{op:hf2}
\end{eqnarray}
\end{enumerate} 
There are fewer experiments involving the third generation than the
first two. In particular, top quark never appears in the final state of
any EWPTs. There is no $\nu_\tau$-nucleon or $\nu_\tau$-lepton
scattering experiment either. Consequently, there exist ``flat 
directions'', that is, combinations of operators that do not
contribute to EWPOs and therefore are not constrained. These flat
directions are expressed as the
relations between the operator coefficients:
\begin{eqnarray}
a_{lQ}^s=-a_{lQ}^t,\quad a_{Le}=-a_{l\tau},\nonumber\\
a_{lL}^s=-a_{lL}^t, \quad a_{hQ}^s=-a_{hQ}^t.\label{flat}
\end{eqnarray}
\subsection{Non-linear case}
If electroweak symmetry is broken by new strong dynamics, the SM model
is described by the non-linear chiral Lagrangian
\begin{equation}
\mathcal{L}=-\frac14B^{\mu\nu}B_{\mu\nu}-\frac14W^{\mu\nu}W_{\mu\nu}+
\frac{v^2}{4}\Tr[(D^\mu U)^\dag (D_\mu U)],\label{eq:nonlinearL}
\end{equation}
where $U$ is a dimensionless unitary unimodular matrix field transform
as a bi-fundamental under $SU(2)_L\times SU(2)_C$. Here $SU(2)_C$
denotes the custodial $SU(2)$. The covariant derivative of $U$ is
defined by
\begin{equation}
D_\mu U=\partial_\mu U+ig\frac{\sigma^a}{2}W^a_\mu U-ig'U\frac{\sigma^3}{2}B_\mu.
\end{equation}
For convenience, we also define
\begin{eqnarray}
&V_\mu= (D_\mu U)U^\dag,\quad T= U\sigma^3U,\nonumber\\
&\hat V_\mu= (D_\mu U^\dag)U,\quad\hat T=\sigma^3.
\end{eqnarray}
Compared with the linear case, the light Higgs is absent from the
Lagrangian. Therefore, the SM predictions are different from those
listed in Table \ref{table:experiments}, but can be estimated by
taking the Higgs mass to a large value in the linear case. To the
leading order, the corrections from a different Higgs mass are given
by\cite{STU} 
\begin{equation}
  S\approx \frac{1}{12 \pi}
  \log\left(\frac{m_h^2}{m_{h,ref}^2}\right), 
  \quad T\approx -\frac{3}{16 \pi c^2}
  \log\left(\frac{m_h^2} {m_{h,ref}^2}\right).
\end{equation}
For more accurate results, one can use the program
GAPP\cite{gapp}, dedicated to the calculations of the SM predictions. 

The difficulty of a strongly coupled electroweak sector is that, the
coefficients of the effective operators below the electroweak symmetry
breaking scale is
generally not calculable. Nevertheless, once we do obtain the
coefficients, which should be small enough to not conflict with the
experiments, the rest of the calculation is analogous to the linear case. 

Assuming CP conservation, we enumerate effective operators in the
chiral Lagrangian, up to dimension-6, and consider their contributions
to the EWPOs. The discussion follows Ref.~\refcite{nonlinear} and
\refcite{disguise}.

There is one dimension-2 operator in addition to the kinetic term of
$U$ in (\ref{eq:nonlinearL}):
\begin{equation}
 {\mathcal L}'_1 =\frac{v^2}{4}\beta_1 [\Tr(TV_\mu)]^2. 
\end{equation}
Going to unitary gauge by setting $U=\mbox{diag}(1,1)$, we obtain a
shift to the $Z$ mass
\begin{equation}
\Delta M_Z^2=-\frac12\beta_1v^2(g^2+{g'}^2).
\end{equation}
Therefore, $\beta_1$ is equivalent to the $T$ parameter:
\begin{equation}
\alpha T=2\beta_1.
\end{equation}
At dimension-4, there are 11 independent operators involving only
gauge bosons and $U$:
\begin{eqnarray}
  {\mathcal L}_1 &=& \frac{1}{2}\alpha_1 g g' B_{\mu \nu} \Tr(W^{\mu \nu} T),\nonumber  \\
  {\mathcal L}_2 &=& \frac{i}{2}\alpha_2 g' B^{\mu \nu} \Tr(T[V_\mu,V_\nu]),\nonumber  \\
  {\mathcal L}_3 &=& i\alpha_3 g \Tr(W_{\mu \nu} [V^\mu,V^\nu]), \nonumber \\
  {\mathcal L}_4 &=& \alpha_4\left[\Tr(V_\mu V_\nu)\right]^2, \nonumber \\
  {\mathcal L}_5 &=& \alpha_5\left[\Tr(V_\mu V^\mu)\right]^2,\nonumber \\
  {\mathcal L}_6 &=& \alpha_6 \Tr(V_\mu V_\nu) \Tr(TV^\mu )  \Tr(TV^\nu ), \nonumber \\
  {\mathcal L}_7 &=& \alpha_7\Tr(V_\mu V^\mu) \Tr(TV^\nu )  \Tr(TV_\nu ),\nonumber  \\
  {\mathcal L}_8 &=& \alpha_8\frac{1}{4} g^2 \left[\Tr(TW_{\mu\nu} )\right]^2,\nonumber  \\
  {\mathcal L}_9 &=& \frac{i}{2}\alpha_9  g \Tr(TW_{\mu\nu}) \Tr([V^\mu,V^\nu] T),  \nonumber \\
  {\mathcal L}_{10} &=& \frac{1}{2} \alpha_{10} \left[\Tr(TV^\mu ) \Tr(TV_\mu  )  \right]^2,\nonumber \\
  {\mathcal L}_{11} &=& \alpha_{11} g \epsilon^{\mu \nu \rho \lambda} \Tr(TV_\mu) \Tr(V_\nu W_{\rho \lambda}).
\end{eqnarray}

These operators can modify the gauge propagators, as well as gauge
boson self couplings. We identify the operators corresponding to $S$
and $U$:
\begin{eqnarray}
S=-16\pi\alpha_1,\quad U=-16\pi\alpha_8.
\end{eqnarray}
Unlike the linear case, the $U$ operator arises at the same order as
the $S$ parameter. 

The triple gauge couplings defined in Eq.~(\ref{eq:triple}) are
modified as
\begin{eqnarray}
g_1^Z-1&=&\frac{1}{c^2-s^2}\beta_1+\frac{1}{c^2(c^2-s^2)}e^2\alpha_1+
\frac1{s^2c^2}e^2\alpha_3,\nonumber\\
g_1^\gamma-1&=&0,\nonumber\\
\kappa_Z-1&=&\frac1{c^2-s^2}\beta_1+\frac1{c^2(c^2-s^2)}e^2\alpha_1+
\frac1{c^2}e^2(\alpha_1-\alpha_2)+\frac1{s^2}e^2(\alpha_3-\alpha_8+\alpha_9),\nonumber\\
\kappa_\gamma-1&=&\frac1{s^2}e^2(-\alpha_1+\alpha_2+\alpha_3-\alpha_8+\alpha_9),\nonumber\\
g_5^Z&=&\frac1{s^2c^2}e^2\alpha_{11},\quad g_5^\gamma=0.
\end{eqnarray}

There are 6 dimension-4 operators containing two fermions:
\begin{eqnarray}
   {\mathcal L}^1_f &=&i \overline{f}_L \gamma^\mu V_\mu f_L,\nonumber \\
   {\mathcal L}^2_f &=&i \overline{f}_L \gamma^\mu (V_\mu T + T V_\mu)
   f_L,\nonumber \\
   {\mathcal L}^3_f &=&i \overline{f}_L \gamma^\mu  T V_\mu T f_L, \nonumber\\
   {\mathcal L}^4_f &=&i \overline{f}_R \gamma^\mu \hat{V}_\mu f_R, \nonumber\\
   {\mathcal L}^5_f &=&i \overline{f}_R \gamma^\mu ( \hat{V}_\mu
   \hat{T} + \hat{T} \hat{V}_\mu) f_R, \nonumber\\
   {\mathcal L}^6_f &=&i \overline{f}_R \gamma^\mu \hat{T} \hat{V}_\mu
   \hat{T} f_R,
\end{eqnarray}
where $f=l,q$. Obviously, these operators affect gauge-fermion
couplings, analogous to
the operators in Eq.~(\ref{eq:ohf}) for the linear case. 
\subsection{Light states and loop corrections}
Due to the stringent constraints shown in Table \ref{table:lambdas},
if the new physical states contribute to EWPOs at tree level, the
constraints on their masses are usually more than a TeV. In this kind
of models, there is a clear energy scale that separates the SM fields
and the heavy states and therefore an effective theory below that
scale is well defined. This is where
the effective theory method finds their best use for.

There are also a
variety of models that new physics contributions to EWPOs do not appear
at tree level. The most extensively studied minimal
supersymmetric standard model (MSSM) is one example. In recent
years, the list has been extended to include universal extra
dimensions with KK parity\cite{ued}, little Higgs models with T
parity\cite{lht} and so 
forth. In these models, interactions involving new particles must
include even number of such particles and therefore only contribute to
EWPOs at loop levels. The loop factor suppression makes it possible to
have very light new particles comparable to the $Z$ mass, and still
pass the electroweak precision
constraints. 

Although stringent constraints from EWPTs have been
avoided, sometime one is interested in how light the new states are
allowed. In principle, the effective theory breaks down
because the new physics does not decouple from the SM. Nevertheless,
since the EWPTs do not have new particles in the initial or final states,
the loop corrections can often be written in the form of effective
operators. In some cases, there are a few such operators dominating the
loop corrections, which greatly simplifies the calculations. We
have already seen an example: the corrections of a different Higgs
mass can be absorbed to the $S$ and $T$ parameters.  As a more complicated
example, in Ref.~\refcite{susybounds}, it is shown that the major
corrections for certain supersymmetric models can be encoded in the oblique
parameters $S$, $T$, $W$ and $Y$. For another examples, in
Ref.~\refcite{lhtbounds}, the loop corrections to the $S$, $T$, $U$
and $\Delta g_L^b$ parameters and 4-fermion operators are calculated for the
littlest Higgs model with T-parity.   
\section*{Acknowledgments}
This work was supported in part by U.S. Department of Energy grant No.
DE-FG03-91ER40674.

\end{document}